\documentclass[12pt, a4paper]{article}

\pdfoutput=1

\usepackage[utf8]{inputenc} 
\usepackage[T1]{fontenc}

\usepackage{amsmath}
\allowdisplaybreaks
\usepackage{amsfonts}
\usepackage{amssymb}
\usepackage{bbm}
\usepackage{verbatim}
\usepackage{dsfont}
\usepackage{booktabs}
\usepackage{slashed}
\usepackage{bm}
\usepackage{subcaption}
\usepackage{cancel}

\usepackage{epsfig}
\usepackage{color}
\usepackage[table]{xcolor}
\usepackage{graphicx}
\usepackage[]{caption}
\usepackage{float}
\usepackage{overpic}

\usepackage{enumerate}
\usepackage{hhline}
\usepackage{multirow}

\usepackage{cite}
\usepackage{xspace}
\usepackage{setspace}

\usepackage{bbold}

\usepackage[title]{appendix}

\renewcommand{\theequation}{\arabic{section}.\arabic{equation}}

\usepackage{ulem}

\newcommand{\V}{{\cal V}}
\newcommand{\M}{{\cal M}}

\newcommand{\half}{\frac{1}{2}}
\newcommand{\dd}{\text{d}}
\newcommand{\overbar}[1]{\mkern 1.5mu\overline{\mkern-1.5mu#1\mkern-1.5mu}\mkern 1.5mu}

\usepackage[pdftitle={Brane supersymmetry breaking without tadpoles (or  Why Supersymmetry is broken)  },
  pdfauthor={},
  pdfsubject={},
  bookmarksopen, bookmarksnumbered, bookmarksopenlevel=2, colorlinks=false, linkcolor=blue, citecolor=blue, urlcolor=blue]{hyperref}

\newcommand{\SO}[1]{\text{SO(\ensuremath{#1})}\xspace}

\renewcommand{\i}{\text{i}}

\renewcommand{\tfrac}{\genfrac{}{}{}1}

\newcommand{\be}{\begin{equation}}
\newcommand{\ee}{\end{equation}}
\newcommand{\ba}{\begin{eqnarray}}
\newcommand{\ea}{\end{eqnarray}}

\begin{document}

\thispagestyle{empty}

\begin{flushright}
CPHT-RR026.032021\\
\end{flushright}
\vskip .8 cm
\begin{center}
  {\Large {\bf Geometry of orientifold vacua and \\supersymmetry breaking }}\\[12pt]

\bigskip
\bigskip 
{\bf Thibaut Coudarchet,\footnote{thibaut.coudarchet@polytechnique.edu}  
Emilian Dudas\footnote{emilian.dudas@polytechnique.edu} and 
Herv\'e Partouche}\footnote{herve.partouche@polytechnique.edu}
\bigskip\\[0pt]
\vspace{0.23cm}
{\it CPHT, CNRS, Ecole Polytechnique, IP Paris, \\F-91128 Palaiseau, France}\\[20pt] 
\bigskip
\end{center}

\date{\today{}}

\begin{abstract}
\noindent

Starting from a peculiar orientifold projection proposed long ago by Angelantonj and Cardella, we elaborate on a novel perturbative scenario that involves  only D-branes, together with the two types of orientifold planes ${\rm O}_{\pm}$ and anti-orientifold planes $\overbar{{\rm O}}_{\pm}$.
We elucidate the microscopic ingredients of such models, connecting them to a novel realization of brane supersymmetry breaking.  Depending on the position of the D-branes in the internal space,
supersymmetry can be broken at the string scale on branes, or alternatively only at the massive level. 
The main novelty of this construction is that it features  no NS-NS disk tadpoles, while avoiding open-string instabilities. 
The one-loop potential, which  depends on the positions of the D-branes, is minimized for maximally broken, nonlinearly realized supersymmetry. 
The orientifold projection and the effective field theory description reveal a soft breaking of 
supersymmetry in the closed-string sector. 
In such models it is possible to decouple the gravitino mass from the value of the scalar potential, while avoiding
brane instabilities. 

\end{abstract}

\newpage 
\setcounter{page}{2}
\setcounter{footnote}{0}

\renewcommand{\baselinestretch}{1.5}

\section{Introduction}
\label{sec:Introduction}

Supersymmetry breaking in string theory is notoriously difficult to achieve in a controllable manner.  There are several challenging and well-known problems to overcome at the string  level
and at the effective field theory  one. 

A generic issue, both at string perturbative and effective supergravity  levels, is that supersymmetry breaking generates potentials for some  moduli fields that are of runaway type, which typically 
drive the dynamics towards zero or strong  string coupling, and also lead to decompactification or compactification of the internal space \cite{Dine:1985he}. The state of the art is to generate a local minimum somewhere far from the runaway 
regime, that is computationally reliable and such that the corresponding lifetime is beyond the age of the universe.
Such a minimum is very hard to obtain in string perturbation theory, and 
easier in practice to obtain at the effective field-theory level, adding extra ingredients like fluxes or nonperturbative effects. 

At the string perturbative level, supersymmetry breaking generates a vacuum energy (more precisely, a scalar potential) at some order in perturbation theory. In
the first models of supersymmetry breaking, so-called Scherk--Schwarz or breaking by compactification \cite{SS,SSstring,SSstring2,SSstring3,SSstring4,SSstring5}, this arises at one loop.\footnote{Scherk--Schwarz compactifications also have often additional, tachyonic-like instabilities in some range of parameters. Tachyon-free examples however exist, see {\it e.g.} \cite{sstachyonfree,sstachyonfree2,sstachyonfree3}.} The generated scalar potential is typically of runaway type\footnote{The cosmological evolution of the moduli fields can be studied in a thermal \cite{runaway-thermal,runaway-thermal2,runaway-thermal3,runaway-thermal4,runaway-thermal5} or cold \cite{runaway-cold,runaway-cold2} universe.} and the classical vacuum used in perturbation theory is therefore not valid anymore. It is however possible, in a more refined construction, to stabilize the corresponding modulus, yielding a negative scalar potential \cite{sstachyonfree2}. 
In a subclass of models, which satisfy a classical Bose/Fermi degeneracy at the massless level, this one-loop potential turns out to be exponentially suppressed at low supersymmetry breaking scale  
\cite{ADM,ADM2,ADM3,ADM4,ADM5,ADM6,ADM7,ADM8,ADM9,ADM10,ADM11,ADM12,ADM13}, but not vanishing \cite{Lnot0}. 

Later on, tachyon-free orientifold string models where supersymmetry is broken at the string scale in the open-string (gauge) sector, whereas the closed-string (gravity) sector is supersymmetric at lowest
perturbative order were constructed \cite{bsb,bsb2,bsb3,bsb4,bsb5,bsb6}.  Since in such frameworks a massless gravitino is present, supersymmetry has to be nonlinearly realized in the open sector and this was indeed
shown explicitly in  \cite{dm,dm2}.  Such models contain non-BPS tachyon-free  configurations. An important application of such setups is the KKLT scenario of moduli stabilization \cite{kklt}, see {\it e.g.} \cite{antibranekklt,antibranekklt2,antibranekklt3,antibranekklt4,antibranekklt5,antibranekklt6,antibranekklt7}. 
There was also a recent simplification in constructing supergravity models with nonlinear supersymmetry \cite{nonlinear,nonlinear2,nonlinear3,nonlinear3bis,nonlinear4}, stimulated in part by ``Brane Supersymmetry Breaking'' (BSB) type models. In such settings,  there is a runaway scalar potential generated at the disk level. Ignoring  the true vacuum state and working naively at fixed values of moduli fields leads to so-called NS-NS tadpoles for 
the corresponding moduli,  which ruin perturbation theory since they generate unphysical UV divergences.  It is widely believed that this does not signal any inconsistency 
of the theory, but just the fact that naive perturbation theory is performed around a point in field space that is not an extremum. Indeed, all models of this type  constructed in the literature satisfy all known consistency conditions.  Mechanisms of shifting the vacuum, in analogy with field-theory examples, were proposed in the literature 
\cite{Fischler:1986tb,Fischler:1986tb2,Fischler:1986tb3,Fischler:1986tb4}. However their practical implementation is limited to toy examples or to special models with small tadpoles. Hence, whereas the BSB models are tachyon-free, the presence of NS-NS tadpoles raises the question of the
validity of perturbation theory and the fate of such constructions \cite{Basile:2018irz, Antonelli:2019nar, Basile:2020mpt,Basile:2020xwi}. Let us also mention  that the  coexistence of massless gravitinos and  broken supersymmetry in the open sector  in BSB models is shared by compactifications with internal magnetic fields that break supersymmetry \cite{magnetic,magnetic2,magnetic3,magnetic4}.

In another class of non-supersymmetric models based on type II asymmetric orbifolds or their orientifold descendants \cite{Z1=0,Z1=02,Z1=03,Z1=04,Z1=05,Z1=06,Z1=07}, a classical Bose/Fermi degeneracy valid at any mass level  implies that the potential arises only at two loops  \cite{L2,L22}. In such frameworks, there are no tadpoles at one loop and no need to shift scalar expectation values for describing vacua at this order of perturbation theory. However, stability at one loop of the moduli fields has not been analyzed.

The goal of the present paper is the construction of BSB string vacua without NS-NS disk tadpoles. Recently, it was conjectured that massless gravitinos in string theory with broken supersymmetry implies a breakdown of the effective field theory \cite{Cribiori:2021gbf,Castellano}.  It is clearly of interest to check  this conjecture in explicit string models, by trying to avoid NS-NS tadpoles. In this paper we identify constructions in which the supersymmetry breaking scale in the gauge (open) sector is much higher than in the closed-string sector. This was actually achieved previously in \cite{Angelantonj:2003hr}. However, our construction avoids the  open-string tachyonic instabilities typically present in such constructions. A runaway behaviour for internal radii is  however still present. We show that the limit of vanishing gravitino mass is inconsistent. The existence of such constructions was already anticipated 
in the pioneering paper  \cite{Angelantonj:2004cm} in an algebraic construction
using the tools of the Tor-Vergata school \cite{AScourse,AScourse2,AScourse3,AScourse4,AScourse5,AScourse6,AScourse7,AScourse8,AScourse9}. We provide here the correct geometric interpretation of the eight-dimensional class of models proposed in 
\cite{Angelantonj:2004cm}, which turns out to contain several types of perturbative orientifold and anti-orientifold planes. 
We also point out that the simultaneous presence of orientifold and antiorientifold planes suggests that the closed-string sector is  not exactly supersymmetric at tree-level, but has softly broken supersymmetry. The basic mechanism goes as follows: 

One starts with a supersymmetric orientifold model containing both  O$_-$  (negative tension, negative RR charge) and O$_+$-planes
(positive tension, positive RR charge). A consistent supersymmetry-breaking deformation of the model turns one   O$_-$ - O$_+$ pair into an  ${\overbar {\rm O}_-}$ - ${\overbar {\rm O}_+}$ pair, which is 
mutually BPS but preserves the other half of the supersymmetries compared to the O$_\pm$-planes and D-branes. Since both the initial O$_-$ - O$_+$ pair and its SUSY breaking
avatar  ${\overbar {\rm O}_-}$ - ${\overbar {\rm O}_+}$  have zero total tension and charge, there will be no RR or NS-NS tadpoles generated in the non-supersymmetric case. Depending on where the background
D-branes sit in the internal space, their massless spectrum can be supersymmetric (if they sit on top of   O$_-$ or O$_+$-planes or in the bulk) or non-supersymmetric  
 (if they sit on top of ${\overbar {\rm O}_-}$  or ${\overbar {\rm O}_+}$-planes, in which case supersymmetry is nonlinearly realized in their worldvolume). Such models also have a supersymmetric limit, when
 a certain radius is taken to zero.\footnote{A similar option is available in IIB flux compactifications\cite{jihad-augusto-unpublished}. We thank J. Mourad and A. Sagnotti for sharing their results with us.} For small values of this radius, the breaking can be interpreted as spontaneous, whereas for large values, supersymmetry breaking can be considered as nonlinearly
 realized if branes sit on top of anti-orientifold planes. Interestingly, naive energetic considerations on brane-orientifold plane interactions suggest that the branes move towards stable configurations 
 with maximal (string scale) breaking of supersymmetry.
 Whereas at first sight the closed-string spectrum could be supersymmetric, we show that a detailed look at the orientifold projections leading to the geometry of O-planes and, independently, considerations from low-energy effective field theory suggest that the correct option is a specific soft supersymmetry breaking deformation in the closed-string sector. A more detailed analysis of the effective field theory of this class of models deserves however a dedicated study. 

The structure of the paper is the following: In Sect.~\ref{sec:8d}, we review the 8d $\text{USp}(16)$ supersymmetric orientifold theory and introduce the novel Brane Supersymmetry Breaking (BSB) mechanism. In particular, we discuss the consistency between the soft breaking of supersymmetry in the closed-string spectrum and the supersymmetry breaking deformation in the open-string  sector. The generalization of the construction to dimensions lower than 8  turns out to be rich but straightforward. We give various examples in Sect.~\ref{sec:lower}. Sect.~\ref{sec:consistency} discusses consistency conditions coming from probe branes as well as nonperturbative constraints to be satisfied by these models. 
In Sect.~\ref{sec:massscales}, we study the supersymmetry breaking mass scales in the closed- and open-string sectors for different positions of stacks of D7-branes in 8d. We also comment on the limit of vanishing gravitino mass and the connexion with the gravitino mass conjecture put forward recently in \cite{Cribiori:2021gbf,Castellano}, in the context of the swampland program \cite{swampland,swampland2,swampland3}. Conclusions and outlooks can be found in Sect.{\ref{sec:conclusion}}, whereas an appendix contains examples of consistent and inconsistent geometric configurations.

\section{The 8d USp(16) superstring and its SUSY breaking avatar}
\label{sec:8d}

In this section, we review the construction of the supersymmetric USp(16) orientifold model in 8 dimensions and then present its non-supersymmetric version.

\subsection{The type IIB torus amplitudes}
\label{sec:8d1}

Let us start by describing alternative viewpoints for deriving the supersymmetric and non-supersymmetric torus amplitudes to be combined later on with orientifold amplitudes.

The original orientifold models described  in  \cite{Bianchi:1991eu,Bianchi:1991eu2}  make
use of a non-trivial quantized background  for the internal components of the antisymmetric tensor field, $B_{ij}$.  This field is odd
under worldsheet parity  and therefore it is projected out by the
orientifold projection $\Omega$ in type I superstring. However, this still leaves
the possibility to add a quantized value
$\frac{2}{\alpha'} B_{ij} \in \mathbb{Z}$, where $\alpha'$ is the string tension. 
In this case, the left and right momenta of closed-string states are given, for a torus factorized into two circles, by
\be
p_{\rm L,R}^{8} = \frac{m_8+ n_9/2}{R_8} \pm \frac{n_8 R_8}{\alpha'} \ ,    \qquad  p_{\rm L,R}^{9} = \frac{m_9- n_8/2}{R_9} \pm \frac{n_9 R_9}{\alpha'} \ . \label{8d01}
\ee
The type IIB torus amplitude is given by 
\begin{align}
{\mathcal T} = \int \frac{\dd^2 \tau}{\tau_2^5}\!\!&\; \left[   \Lambda_{m_9,2n_9} \Lambda_{m_8,2n_8} +   \Lambda_{m_9+1/2, 2n_9} \Lambda_{m_8, 2n_8+1}    \right. \nonumber \\
&\! \left. + \,\Lambda_{m_9, 2n_9+1} \Lambda_{m_8 +1/2, 2n_8}
+   \Lambda_{m_9+1/2, 2n_9+1} \Lambda_{m_8+1/2, 2n_8+1}   \right]  \left| \frac{V_8-S_8}{\eta^8}\right|^2 \ , \label{8d1} 
\end{align}
where $V_8$, $S_8$ (along with $O_8$, $C_8$) are the ${\rm SO}(8)$ affine characters and  $\eta$ is the Dedeking function. They all depend on the Teichm\"uller parameter $\tau$ of the genus-1 surface, whose imaginary part is denoted $\tau_2$. Moreover, the lattices are expressed in terms of 
\be
\Lambda_{m_i,n_i} = q^{{\alpha'\over 4}\big({m_i\over R_i}+n_i{R_i\over \alpha'}\big)^2}\, \bar q^{{{\alpha'}\over 4}\big({m_i\over R_i}-n_i{R_i\over \alpha'}\big)^2}\ , \qquad q=e^{2i\pi \tau}\ ,
\ee
where $m_i$, $n_i$ are the momentum and winding numbers along direction $X^i$.\footnote{Throughout our work, all discrete sums over integer $m_i$, $n_i$ are  implicit. The conventions used in partition functions are those given \textit{e.g.} in the reviews \cite{AScourse8,AScourse9}.} Note that this amplitude is invariant under the T-duality transformation $(R_8,R_9)\to \big(\frac{\alpha'}{2 R_8}, \frac{\alpha'}{2 R_9}\big)$.

There is another particularly useful way of constructing the torus amplitude with non-trivial discrete antisymmetric tensor uncovered by Pradisi \cite{Pradisi:2003ct}. The starting point  is a freely-acting orbifold of type IIB with  $B_{89}=0$ and generator $g=\delta_{w_8}\delta_{p_9}$, where $\delta_{w_8}$ stands for a winding shift along direction $X^8$ while $\delta_{p_9}$ denotes a momentum shift along direction $X^9$. The action of this generator on the lattice states is
\begin{equation}
g|{\bf m} , {\bf n} \rangle = (-1)^{n_8+m_9}|{\bf m} , {\bf n} \rangle\ .
\end{equation}
The gauging of the theory with this generator implies the existence of four contributions in the torus amplitude corresponding to the untwisted and twisted sectors, both with or without insertion of the orbifold generator in the traces. One obtains
\begin{equation}
\mathcal{T}=\half\int\frac{\dd^2\tau}{\tau_2^5}\!\!\;\left[1+(-1)^{n_8+m_9}\right]\left(\Lambda_{m_8,n_8}\Lambda_{m_9,n_9}+\Lambda_{m_8+\half,n_8}\Lambda_{m_9,n_9+\half}\right)\left| \frac{V_8-S_8}{\eta^8}\right|^2 \ .
\end{equation}
A rescaling of the radius $R_9\to 2R_9$ then leads to the torus amplitude with discrete antisymmetric tensor given in  Eq.~(\ref{8d1}).

Note that another derivation can be obtained by applying the T-duality transformation $R_8\to {\alpha'\over R_8}=\tilde R_8$ on the freely-acting orbifold of type IIB with $B_{89}=0$. In fact, the complex coordinate $Z = \frac{\tilde X_8+ i X_9}{2 \pi \tilde R_8}$, where $\tilde X_8$ is the T-dual coordinate, satisfies the identifications $Z = Z+1 = Z + i \frac{R_9}{\tilde R_8}$. Moreover, the orbifold generator $g=\delta_{w_8}\delta_{p_9}$ is mapped to ${\tilde g}=\delta_{p_8}\delta_{p_9}$
defined as $(\tilde X_8,X_9)= (\tilde X_8 + \pi \tilde R_8, X_9 + \pi R_9)$. 
We have therefore three identifications, which can be encoded in the following two:
\begin{equation}
Z = Z+ 1 \ , \quad Z = Z + U \ ,
\quad {\rm where} \quad  U = \frac{1}{2} + i \frac{R_9}{2 \tilde R_8} \ . 
\end{equation}
Hence, by rescaling $R_9 \to 2 R_9$, the coordinate $Z$ is that of a tilted torus of complex structure 
$U = \frac{1}{2} + i \frac{R_9}{\tilde R_8}$. However, it is known that the type IIA theory compactified on this tilted torus is T-dual to the type IIB theory with antisymmetric background $B_{89}={\alpha'\over 2}$. 

From the freely-acting orbifold perspective, it is now easy to build a non-super\-symmetric deformation of the type IIB model in a Scherk--Schwarz spirit. It is obtained by replacing $g$ with the generator $g'=(-1)^F\delta_{w_8}\delta_{p_9}$, where $F$ denotes the spacetime fermion number. The construction of the torus amplitude is straightforward and the result is
\begin{align}
{\mathcal T} = &\; {1\over 2}\int \frac{\dd^2 \tau}{\tau_2^5}\!\!\; \bigg\{   \Lambda_{m_8,n_8}\Lambda_{m_9,n_9}\left|V_8-S_8\right|^2 + (-1)^{n_8+m_9}\Lambda_{m_8,n_8}\Lambda_{m_9,n_9}\left|V_8+S_8\right|^2\nonumber \\
&+\Lambda_{m_8+\half,n_8}\Lambda_{m_9,n_9+\half}\left|O_8-C_8\right|^2+(-1)^{n_8+m_9}\Lambda_{m_8+\half,n_8}\Lambda_{m_9,n_9+\half}\left|O_8+C_8\right|^2\bigg\}\frac{1}{|\eta^8|^2}\ . 
\end{align}
The rescaling of the radius $R_9\to 2R_9$ then leads to
\begin{align}
{\mathcal T} = \int \frac{\dd^2 \tau}{\tau_2^5}&\!\!\; \bigg\{\;\left(\Lambda_{m_8,2n_8} \Lambda_{m_9,2n_9}+\Lambda_{m_8,2n_8+1} \Lambda_{m_9+\half,2n_9}\right)\left(|V_8|^2+|S_8|^2\right)\nonumber\\
&-\left(\Lambda_{m_8,2n_8+1} \Lambda_{m_9,2n_9}+\Lambda_{m_8,2n_8} \Lambda_{m_9+\half,2n_9}\right)\left(V_8\overline{S}_8+\overline{V}_8S_8\right)  \nonumber\\
&+\left(\Lambda_{m_8+\half,2n_8} \Lambda_{m_9,2n_9+1}+\Lambda_{m_8+\half,2n_8+1} \Lambda_{m_9+\half,2n_9+1}\right)\left(|O_8|^2+|C_8|^2\right)\nonumber\\
&-\left(\Lambda_{m_8+\half,2n_8+1} \Lambda_{m_9,2n_9+1}+\Lambda_{m_8+\half,2n_8} \Lambda_{m_9+\half,2n_9+1}\right)\left(O_8\overline{C}_8+\overline{O}_8C_8\right)\bigg\}\frac{1}{|\eta^8|^2}\ .\label{torus2}
\end{align}
The type IIB gravitinos acquire masses 
\begin{equation}
M_1=\frac{R_8}{\alpha'} \qquad\text{or}\qquad M_2=\frac{1}{2R_9} \ , 
\label{grav-mass}
\end{equation}
which vanish  in the supersymmetric limits $R_8\to 0$ and/or $R_9\to\infty$. Moreover, as usual with a Scherk--Schwarz mechanism, a scalar in the twisted sector becomes tachyonic when the radii satisfy
\begin{equation}
\frac{1}{4R_8^2}+\frac{R_9^2}{\alpha'^2}<\frac{2}{\alpha'}\ .
\end{equation}

Notice that, unlike its supersymmetric version 
(\ref{8d1}), the  torus amplitude 
(\ref{torus2}) is not invariant  under the T-duality $(R_8,R_9)\to \big(\frac{\alpha'}{2 R_8}, \frac{\alpha'}{2 R_9}\big)$. Indeed, this transformation amounts to 
exchanging the lattice sums of the two directions and thus switching \mbox{$X^8 \leftrightarrow X^9$}, leading to the new amplitude 
\begin{align}
{\tilde {\mathcal T}} = \int \frac{\dd^2 \tau}{\tau_2^5}&\!\!\; \bigg\{\;\left(\Lambda_{m_8,2n_8} \Lambda_{m_9,2n_9}+ \Lambda_{m_8+\half,2n_8} \Lambda_{m_9,2n_9+1} \right)\left(|V_8|^2+|S_8|^2\right)\nonumber\\
&-\left(\Lambda_{m_8,2n_8}  \Lambda_{m_9,2n_9+1} + \Lambda_{m_8+\half,2n_8} \Lambda_{m_9,2n_9} \right)\left(V_8\overline{S}_8+\overline{V}_8S_8\right) \nonumber \\
&+\left( \Lambda_{m_8,2n_8+1} \Lambda_{m_9+\half,2n_9} + \Lambda_{m_8+\half,2n_8+1} \Lambda_{m_9+\half,2n_9+1} \right)\left(|O_8|^2+|C_8|^2\right)\nonumber\\
&-\left( \Lambda_{m_8,2n_8+1} \Lambda_{m_9+\half,2n_9+1} +\Lambda_{m_8+\half,2n_8+1} \Lambda_{m_9+\half,2n_9} \right)\left(O_8\overline{C}_8+\overline{O}_8C_8\right)\bigg\}\frac{1}{|\eta^8|^2}\label{torus3}\ . 
\end{align}
The latter can therefore be obtained  by a free action generated by $g'' = (-1)^F \delta_{p_8}\delta_{w_9}$, followed by the rescaling $R_8 \to 2 R_8$.  In this case, the masses of the gravitinos are
\begin{equation}
 M_1=\frac{1}{2R_8} \qquad\text{or}\qquad M_2=\frac{R_9}{\alpha'}\label{grav-mass2}
\end{equation}
and supersymmetry is restored in the limits $R_8 \to \infty$ and/or $R_9 \to 0$. 

\subsection{The supersymmetric orientifold amplitudes}
\label{sec:8d2}

In  eight dimensions, the gauge group in supersymmetric orientifold models has rank $16$, $8$ or 0.\footnote{We refer only to the gauge group arising from  the open-string/D-brane sector.} For rank $8$, the gauge group of maximal dimension, {\it i.e.} in the absence of Wilson lines,  is $\text{USp}(16)$. This 8d model  was first constructed by Bianchi, Pradisi and
Sagnotti in terms of D9-branes and an O$9_-$-plane \cite{Bianchi:1991eu,Bianchi:1991eu2}. It also  has  a dual description in terms of CHL strings \cite{Chaudhuri:1995fk}. Moreover, 
it admits a geometrical T-dual picture  understood later on  by Witten, which we will consider hereafter \cite{Witten:1997bs}.  

\begin{figure}[h!]
\captionsetup[subfigure]{position=t}
\begin{subfigure}[t]{0.31\textwidth}
\begin{center}
\includegraphics[scale=0.6]{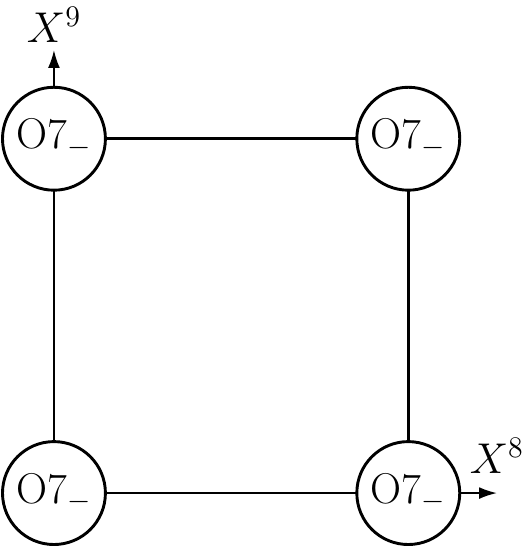}
\end{center}
\caption{\footnotesize Geometry of the standard $\text{SO}(32)$ superstring. There is an ${\rm O}7_-$-plane at each of the four fixed points.}
\label{8d-square1}
\end{subfigure}
\quad
\begin{subfigure}[t]{0.31\textwidth}
\begin{center}
\includegraphics[scale=0.6]{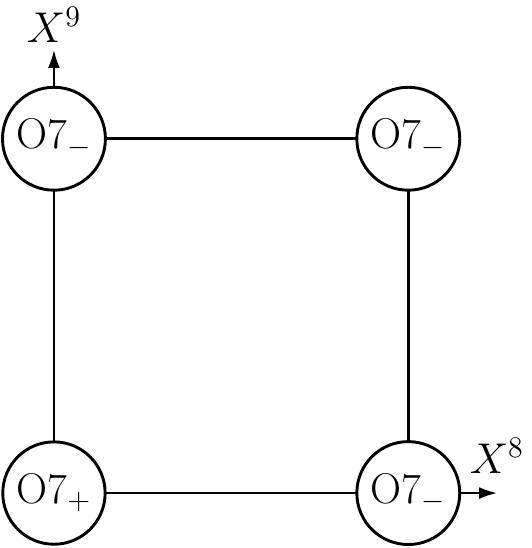}
\end{center}
\caption{\footnotesize Geometry of the supersymmetric $\text{USp}(16)$ model. There is 1 ${\rm O}7_+$-plane at $(0,0)$ and 3 ${\rm O}7_-$-planes at $(\pi R_8,0)$, $(0,\pi R_9)$ and $(\pi R_8,\pi R_9)$.}
\label{8d-square2}
\end{subfigure}
\quad
\begin{subfigure}[t]{0.31\textwidth}
\begin{center}
\includegraphics[scale=0.6]{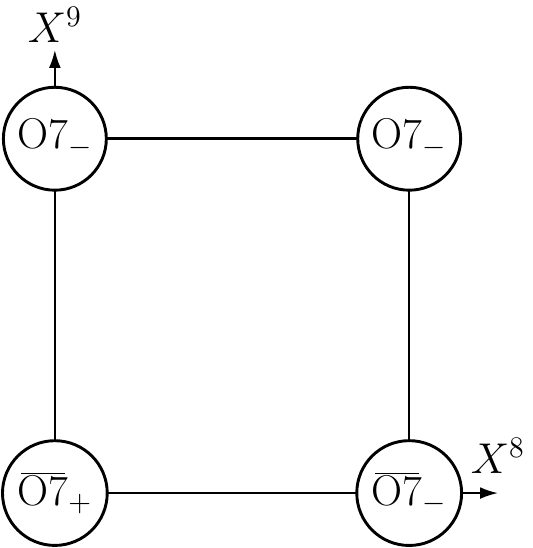}
\end{center}
\caption{\footnotesize Geometry of the non-supersymmetric $\text{USp}(16)$ model. There is 1 $\overbar{{\rm O}7}_+$-plane at $(0,0)$, 1 $\overbar{{\rm O}7}_-$-plane at $(\pi R_8,0)$ and 2 ${\rm O}7_-$-planes at $(0,\pi R_9)$ and $(\pi R_8,\pi R_9)$.}
\label{8d-square3}
\end{subfigure}
\caption{\footnotesize Eight dimensional T-dual geometries: The standard $\text{SO}(32)$ superstring theory, the $\text{USp}(16)$ supersymmetric theory and its non-supersymmetric version.}
\end{figure}

In the case of the standard $\text{SO}(32)$ type I superstring, {\it i.e.} with $B_{89}=0$, the standard T-duality transformation $(R_8,R_9)\to ({\alpha'\over R_8},{\alpha'\over R_9})$ turns the $\text{O}9_-$-plane wrapping the torus into four
$\text{O}7_-$-planes located at the orientifold fixed points of the generator \mbox{$\Omega' = \Omega \Pi_8 \Pi_9 (-1)^{F_{\rm L}}$}. In our notations, $\Pi_i$ is a parity operation $X^i\to - X^i$ and $F_{\rm L}$ is the left-handed  fermion number. This geometry is depicted in Fig.~\ref{8d-square1} and the  resulting model contains 16 D7-branes\footnote{Or equivalently 32 ``half-branes'' of type IIB organized as 16 mirror pairs referred to as 16 ``branes'' in the orientifold theory.} in order to  cancel the RR tadpole.

On the other hand, it has been shown that for the $\text{USp}(16)$ theory, {\it i.e.} with $B_{89}={\alpha'\over 2}$, the T-duality transformation $(R_8,R_9)\to ({\alpha'\over 2R_8},{\alpha'\over 2R_9})$ turns the original $\text{O}9_-$-plane into three $\text{O}7_-$-planes and  one $\text{O}7_+$-plane \cite{Witten:1997bs}. While an $\text{O}7_-$-plane has charge (and tension) equal to $-4$ in units of a regular D7-brane charge,  an $\text{O}7_+$-plane has charge (and tension)  equal to $+4$. The geometry is depicted in Fig.~\ref{8d-square2}, where $R_8$, $R_9$ now refer to the radii in the T-dual theory. The switch $\text{O}7_- \to \text{O}7_+$ has the overall effect of halving the RR tadpole,  a fact that requires the addition of only eight D7-branes (16 half-branes). The rank of the gauge group is thus reduced to 8. Furthermore, while  D7-branes on top of an $\text{O}7_-$-plane lead to an orthogonal (SO) gauge group, D7-branes on top of an $\text{O}7_+$-plane lead to a symplectic (USp) gauge factor. Therefore, the configuration with all the D7-branes sitting on top of the $\text{O}7_+$-plane yields the gauge symmetry $\text{USp}(16)$. 

From now on, the description of the  $\text{USp}(16)$ theory we choose is that of the type IIB theory with orientifold projection  $\Omega' = \Omega \Pi_8 \Pi_9 (-1)^{F_{\rm L}}$, which involves $\text{O}7_\pm$-planes and D7-branes. The spectrum is encoded in the partition functions which can be worked out using standard methods.  The torus contribution is given by half that given in Eq.~({\ref{8d1}).\footnote{Remind  that the type IIB amplitude (\ref{8d1}) is self-dual. Hence, it can be used in the orientifold theory obtained by  modding with $\Omega$ \cite{Bianchi:1991eu,Bianchi:1991eu2} or that obtained with $\Omega'$. Being T-dual, they are physically equivalent.} 
Moreover, the Klein, cylinder and M\"obius amplitudes are 
\begin{align}
\mathcal{K} &= \frac{1}{2} \int_0^{\infty} \frac{\dd \tau_2}{\tau_2^5}  \,W_{2n_9}  W_{2n_8}  \frac{V_8-S_8}{\eta^8} \big(2i \tau_2\big) \ , \nonumber \\
\mathcal{A} &= \frac{N^2}{2} \int_0^{\infty} \frac{\dd \tau_2}{\tau_2^5}\, W_{n_9} W_{n_8}    \frac{V_8-S_8}{\eta^8} \big(\tfrac{i \tau_2}{2}\big)  \ , \nonumber \\
\mathcal{M} &= \frac{N}{2} \int_0^{\infty} \frac{\dd \tau_2}{\tau_2^5}  \, W_{n_9} \big[ (-1)^{n_9}  W_{2n_8} - W_{2n_8+1} \big]  \frac{{\hat V}_8-{\hat S}_8}{{\hat \eta}^8} (\tfrac{i\tau_2}{2}+\tfrac{1}{2}) 
  \  ,   \label{8d2}  
 \end{align}
 where $N$ is the number of half-D7-branes and the lattices of winding modes are defined as 
 \be
W_{n_i} = e^{-\pi \tau_2 n_i^2 \frac{R_i^2} {\alpha'}}  \ .  \label{8d3} 
 \ee 
The ``field-theory'' open-string spectrum is encoded in 
\begin{align}
 (\mathcal{A} + \mathcal{M})|_{\rm FT} = \int_0^{\infty} \frac{\dd \tau_2}{\tau_2^5}
    &\left[ \frac{N(N+1)}{2}
   W_{2n_9}W_{2n_8}  \right. \nonumber \\
  &\left. + \frac{N(N-1)}{2}\big(W_{2n_9+1}W_{2n_8}+
   W_{n_9}W_{2n_8+1} \big) \right] \left.\frac{V_{8}-S_{8}}{\eta^8}\right|_{0} \   ,\label{8d4}
\end{align}
where the index $0$ stands for the constant mode of the characters. It is manifestly supersymmetric and describes a $\text{USp}(N)$ gauge symmetry. The value $N=16$ is found by imposing the RR tadpole condition, which can be derived  from  the amplitudes in the tree-level channel,
\begin{align}
\mathcal{\tilde K} &= \frac{2^5 \alpha'}{8 R_9 R_8 } \int_0^{\infty} \dd l \, P_{m_9}  P_{m_8} \frac{V_8-S_8}{\eta^8} \big(i l\big) \ , \nonumber \\
  \mathcal{\tilde A} &=\frac{2^{-5} N^2 \alpha'}{2 R_9 R_8 } \int_0^{\infty} \dd l \, P_{m_9} P_{m_8}   \frac{V_8-S_8}{\eta^8} \big( il \big) \ , \nonumber \\
\mathcal{\tilde M}& = \frac{N \alpha'}{2 R_9R_8} \int_0^{\infty} \dd l  \, \big[  P_{2m_9+1}  - (-1)^{m_8} P_{2m_9}  \big]  P_{m_8}
\frac{{\hat V}_8-{\hat S}_8}{{\hat \eta}^8} (i l+\tfrac{1}{2})   \  ,   \label{8d5}  
\end{align}
where the lattices of momentum states are given by 
 \be
P_{m_i} = e^{-\pi \frac{l}{2} m_i^2 \frac{\alpha'}{R_i^2}}  \ .  \label{8d3b} 
 \ee 
 
 The tree-level amplitudes encode uniquely the geometry of the D-branes and O-planes.  Indeed, the geometry can in general be  revealed by remembering that the tree-level channel amplitudes capture the propagation of closed strings between orientifold planes and/or  D-branes. As an example, a generic Klein-bottle amplitude can be formally written as
 \be
 \mathcal{\tilde K} = \sum_{a,{\bf m}}\sum_{A,B}(-1)^{F_{\rm L}} C_{a A} \, C_{a B}\,  G_{a{\bf m}} ({\bf x}_A,{\bf x}_B)  \ , \label{8d7} 
 \ee 
 where $a$ labels the NS-NS and RR closed-string degrees of freedom in ten dimensions and ${\bf m} =(m_8,m_9)$ (in 8d models as above) are the internal momenta of their Kaluza--Klein (KK) modes. Moreover, $G_{a{\bf m}} ({\bf x}_A,{\bf x}_B)$ is the tree-level scalar propagator transverse to the O-planes for a flat internal space, while  $C_{a A}$ captures the coupling to the  O-plane $A$ located at ${\bf x}_A= (x^8_A,x^9_A)$. In our examples, $C_{aA}\propto T_A$ for the NS-NS states and $C_{aA}\propto Q_A$ for the RR ones, where $T_A$ and $Q_A$ denote the tension and charge of the O-plane~$A$ and the proportionality constants are equal.
 Actually, the closed-string states $a,{\bf m}$ propagating in $\mathcal{\tilde K}$, which are bosons arising in the NS-NS and RR sector, have different Lorentz structures: For instance the dilaton is a scalar, the graviton is a tensor, etc. Hence, they have different propagators and couplings to branes and orientifolds. Contracting the couplings and the propagators, one obtains the effective couplings $C_{a A}$ and a scalar propagator $G_{a{\bf m}}({\bf x}_A,{\bf x}_B)$ for each closed-string state. Explicitly, we have
 \begin{align}
G_{a{\bf m}}({\bf x}_A,{\bf x}_B) &= e^{i {\bf m} ( {\bf x}_A-{\bf x}_B)} \frac{1}{p_{\parallel}^2 + M_a^2+\sum_i \frac{m_{i}^2}{R_i^2}} \nonumber \\
& = \frac{\pi \alpha'}{2}  e^{i {\bf m} ( {\bf x}_A-{\bf x}_B)}
\int_0^{\infty} \dd l \ e^{- \pi \frac{ l }{2} \alpha'\big(M_a^2+\sum_i  \frac{m_{i}^2}{R_i^2}\big)}   \  ,  \label{8d8} 
\end{align}
where by convention the variables $x^i_A$ take values in the range $[- \pi ,\pi]$ and the internal coordinates are defined as $X^i=x^iR_i$. 
 The closed-string channel Klein-bottle amplitude is therefore given by
\be
\mathcal{\tilde K} =  \frac{\pi \alpha'}{2} \sum_{a,{\bf m}} \sum_{A,B} C_{a A} C_{a B} \int_0^{\infty}  \dd l \ e^{i {\bf m} ( {\bf x}_A-{\bf x}_B)- \pi \frac{ l }{2} \alpha'\big(M_a^2+\sum_i  \frac{m_{i}^2}{R_i^2}\big)}   \  .  \label{8d9} 
 \ee
 The factors $e^{i {\bf m} ( {\bf x}_A-{\bf x}_B)}$ capture the locations of the  O-planes and display the products of the wavefunctions of a closed-string Kaluza--Klein mode $a,{\bf m}$  respectively located on the O-planes $A$ and $B$. 
 
In the 8d examples of this section, there are four orientifold fixed points $(0,0)$, $(0,\pi R_9)$,$(\pi R_8,0)$, $(\pi R_8, \pi R_9)$, where the O7-planes sit. The phases $e^{i {\bf m} ( {\bf x}_A-{\bf x}_B)}$ encoding the propagation between the four O-planes take values $1$, $(-1)^{m_9}$, $(-1)^{m_8}$ or $ (-1)^{m_9+m_8}$. Once dressed by the signs given by the  tensions and charges, they produce projectors in the tree-level channel amplitude. In the $\SO{32}$ type IIB orientifold case, which  contains 4 ${\rm O}7_-$-planes, the projector in the tree-level Klein-bottle amplitude  is
 \be
\Pi_{ \mathcal{\tilde K}} \propto 4 [1 + (-1)^{m_9}]     [1 + (-1)^{m_8}]    \ ,\label{8d08}
 \ee
which projects onto even KK states. In contrast,  the corresponding one for the supersymmetric $\text{USp}(16)$ type IIB orientifold satisfies
\begin{align}
\Pi_{ \mathcal{\tilde K}} &\propto 4  -2  (-1)^{m_9}  +2  (-1)^{m_9} -  2 (-1)^{m_8}  +  2 (-1)^{m_8}   -  2 (-1)^{m_9+m_8}  +  2 (-1)^{m_9+m_8}\nonumber \\
&\propto 4\ ,  \label{8d08bis}
\end{align}
leading to no projection of the KK states.

The tree-level channel cylinder and M\"obius amplitudes take similar formal expressions. In the former case, the objects $A$ and $B$ are D-branes while in the latter case they are a D-brane and an O-plane. For instance, in the ${\rm USp}(16)$ model, the lattices in the  M\"obius amplitude involve all momentum states subject to the projector
 \be
 \Pi = \frac{1-(-1)^{m_9}- (-1)^{m_8} - (-1)^{m_9+m_8}}{2}  \  ,  \label{8d6}  
  \ee
 which is consistent with the geometry of one stack of $8$ (regular) D7-branes coincident with the O$7_+$-plane, whereas the three other orientifold fixed points are occupied by standard  O$7_-$-planes. Moving all the D7-branes on top of one of the O$7_-$-planes lead to an   $\text{SO}(16)$ gauge group, whereas moving all of them into the bulk in one stack leads to a $\text{U}(8)$ open-string gauge group.

\subsection{The non-supersymmetric orientifold amplitudes}
    
Let us now turn to the implementation of supersymmetry breaking in the Klein, cylinder and M\"obius amplitudes, {\it without introducing perturbative instabilities or tadpoles}. The allowed form of the corresponding torus amplitude will be determined in the next subsection.

Geometrically, the mechanism of supersymmetry breaking is the following. A pair of O$7_+$-plane and O$7_-$-plane have globally zero tension and RR charge. From a string-theory  viewpoint, it is possible to replace such a pair by an $\overbar{{\rm O}7}_+$ and $\overbar{ {\rm O}7}_-$ pair, which also has vanishing total tension and charge. However, the second option breaks supersymmetry in the presence of the $8$ D7-branes needed to cancel the tadpoles. This geometry is depicted in Fig.~\ref{8d-square3}. Supersymmetry breaking is not visible in the cylinder amplitude, which describes D7-D7 amplitudes. It is less obvious but true that the orientifold configuration with 2 O$7_-$, 1 $\overbar{ {\rm O}7}_+$, 1 $\overbar{ {\rm O}7}_-$ and the supersymmetric one with 1 O$7_+$ and 3 O$7_-$-planes lead to  identical Klein-bottle amplitudes. 
Indeed, in the former case we have 
\begin{align}
\mathcal{\tilde K} &\propto  \Bigl\{ [4 - 2 (-1)^{m_8}+2 (-1)^{m_8} ] (V_8-S_8) + \nonumber \\
&\ \ \ \ \ \;\, [- 2 (-1)^{m_9}+ 2 (-1)^{m_9} -  2 (-1)^{m_9+m_8}+2 (-1)^{m_9+m_8} ] (V_8+S_8) \Big\}  P_{m_8}  P_{m_9}\nonumber \\
&\propto 4(V_8-S_8)P_{m_8}  P_{m_9}\ ,  \label{8d09}
\end{align}
where the character $V_8-S_8$ describes the tree-level propagation of closed strings between mutually BPS O7-planes (O$7_{-}$-O$7_{-}$,  $\overbar {\rm O7}_{\pm}$-$\overbar {\rm O7}_{\pm}$), whereas  $V_8+S_8$ describes the tree-level propagation of closed strings between mutually non-BPS O7-planes (O$7_{-}$-$\overbar {\rm O7}_{\pm}$, $\overbar {\rm O7}_{\pm}$-O$7_{-}$). The phases 
reflect the geometry of O-planes and lead to a cancellation  of the non-BPS terms, leaving  the unprojected supersymmetric sum over all KK states as in the supersymmetric case (see Eqs.~(\ref{8d5}) and~(\ref{8d08bis})). 

As will be shown later, the configuration where the $8$ D7-branes are coincident with the $\overbar{ {\rm O}7}_+$-plane is the only stable configuration at one loop. Using the above given geometrical interpretation of O-planes, it is easy to check that the tree-level channel amplitudes are given by  
\begin{align}
\mathcal{\tilde K} &= \frac{2^5 \alpha'}{8 R_9 R_8 } \int_0^{\infty} \dd l \, P_{m_9}  P_{m_8} \frac{V_8-S_8}{\eta^8} \big(i l\big) \ , \nonumber \\
\mathcal{\tilde A} &=\frac{2^{-5} N^2 \alpha'}{2 R_9 R_8 } \int_0^{\infty} \dd l \,  P_{m_9} P_{m_8}   \frac{V_8-S_8}{\eta^8} \big( il \big) \ , \nonumber \\
\mathcal{\tilde M} &= \frac{N \alpha'}{2 R_9R_8} \int_0^{\infty} \dd l   \,\big[  P_{2m_9+1}  - (-1)^{m_8} P_{2m_9} \big]  P_{m_8}
\frac{{\hat V}_8- (-1)^{m_8}  {\hat S}_8}{{\hat \eta}^8} (i l+\tfrac{1}{2})   \  .   \label{8d10}  
\end{align}
Notice that the only change in (\ref{8d10}) compared to the supersymmetric case (\ref{8d5})  is the extra phase $(-1)^{m_8}$ in the RR couplings of the closed strings propagating between the D7-branes and the O7-planes in the M\"obius amplitude. The projector in the RR sector is thus transformed accordingly,
\begin{align}
\Pi_{\rm NSNS} &=  \frac{+1-(-1)^{m_9}- (-1)^{m_8} - (-1)^{m_9+m_8}}{2}\nonumber \\
 \Pi_{\rm RR} = (-1)^{m_8}  \Pi_{\rm NSNS} &=  \frac{-1-(-1)^{m_9}+ (-1)^{m_8} - (-1)^{m_9+m_8}}{2}  \  ,  \label{8d11}  
  \end{align}
  where  $\Pi_{\rm NSNS}$ is identical to that of the supersymmetric case, Eq.~({\ref{8d6}). 
The above projectors  encode all the information about the geometry. In fact, with all D-branes located at the origin, the change of signs of the  RR  couplings at $(0,0)$ and $(\pi R_8,0)$ tells us that the orientifold planes located there preserve opposite supersymmetry as compared to the D7-branes and are therefore $\overbar{ {\rm O}7}_+$ and $\overbar{ {\rm O}7}_-$. 

The loop-channel amplitudes can be worked out by the usual methods, leading to 
\ba
&& \mathcal{K} = \frac{1}{2} \int_0^{\infty} \frac{\dd \tau_2}{\tau_2^5}  \, W_{2n_9}  W_{2n_8}  \frac{V_8-S_8}{\eta^8} \big(2i \tau_2\big) \ , \nonumber \\
&&   \mathcal{A} = \frac{N^2}{2} \int_0^{\infty} \frac{\dd \tau_2}{\tau_2^5} \, W_{n_9} W_{n_8}    \frac{V_8-S_8}{\eta^8} \big(\tfrac{i \tau_2}{2}\big)  \ , \nonumber \\
&& \mathcal{M} = \frac{N}{2} \int_0^{\infty} \frac{\dd \tau_2}{\tau_2^5}   \, W_{n_9} \big[ (-1)^{n_9}  W_{2n_8} -  W_{2n_8+1} \big]  \frac{{\hat V}_8 + (-1)^{n_9}{\hat S}_8 }{{\hat \eta}^8} (\tfrac{i\tau_2}{2}+\tfrac{1}{2}) 
   \  .   \label{8d12}  
 \ea
Comparing the M\"obius amplitude with its supersymmetric counterpart (\ref{8d2}) reveals a supersymmetry breaking
orientifold projection
\be
\Omega'' = \Omega \Pi_8 \Pi_9 (-1)^{F_{\rm L}} (-\delta_{w_9})^F
\label{omega}\ ,
\ee
 where, as before, $\Pi_i$ is the parity operation $X^i\to -X^i$, $F_{\rm L}$
is the left-moving fermion number, $F$ is the spacetime fermion number and $\delta_{w_9}$ generates a winding shift in the  coordinate $X^9$. The latter  acts on the zero-modes as $|{\bf m} , {\bf n} \rangle \to (-1)^{n_9}
 |{\bf m} , {\bf n} \rangle$, as follows from a left-right asymmetric action $X_{\rm L}^9 \to X_{\rm L}^9 + \frac{\pi R_9}{2}$,
 $X_{\rm R}^9 \to X_{\rm R}^9 - \frac{\pi R_9}{2}$ on the left- and right-moving parts of the coordinate. Notice that since there is no fermion propagating in the Klein bottle, the supersymmetry breaking deformation $(-\delta_{w_9})^F$ has no effect in this amplitude. 

The massless field-theory open-string spectrum is  captured by 
\be
  (\mathcal{A} + \mathcal{M})|_{\rm FT} = \int_0^{\infty} \frac{\dd \tau_2}{\tau_2^5}
    \left[ \frac{N(N+1)}{2} \left. {V_8\over \eta^8}\right|_0  - \frac{N(N-1)}{2} \left.{S_{8}\over \eta^8}\right|_0 \right] \   \label{8d13}
\ee
and displays supersymmetry breaking at the string scale, of the brane supersymmetry breaking type. The gauge group is $\text{USp}(16)$ as before. The vector bosons are thus in the symmetric representation,
but the fermions are in the antisymmetric representation, which contains in particular the gauge-singlet goldstino. This is the basic indication of the nonlinear realization of  supersymmetry where the D7-branes sit. 
It is then easy to move D7-branes in the internal two-torus and derive the resulting spectrum. Putting all D-branes on top of the $\overbar{ {\rm O}7}_-$-plane  leads to an $\text{SO}(16)$ gauge group with massless fermions in the symmetric representation. The latter contains the singlet goldstino implying again a nonlinear supersymmetry and a supersymmetry breaking at the string scale.  Putting all D7-branes on top of one of the two remaining 
${{\rm O}7_-}$-planes leads to a supersymmetric massless spectrum with $\text{SO}(16)$ gauge group and a supersymmetry breaking at the massive level due to the far-away presence of the two $\overbar{{\rm O}7}$-planes.

  Let us stress again that despite the fact that the O7-planes are of  types orientifold and anti-orientifold, the Klein bottle amplitude
 is exactly the same as in the supersymmetric case, due to the cancellation of the supersymmetry-breaking contributions.\footnote{We will discuss in the next subsection the issue of supersymmetry in the closed-string spectrum. The O-planes and anti O-planes are mutually non-BPS. The cancellation of the non-supersymmetric contributions in the Klein bottle does not  mean that the closed-string sector is supersymmetric, even ignoring the supersymmetry breaking transmission from the open sector. As we will see, the most plausible possibility is that the tree-level closed-string spectrum has softly broken supersymmetry. Another possibility, which we consider however unlikely, is that the closed-string spectrum is supersymmetric but the interactions are not. Another insight about this issue is the gravitino masses: A supersymmetric closed-string spectrum would be in contradiction with the presence of orientifold and anti-orientifold planes, which impose opposite boundary conditions for the gravitinos.}  At one loop, only  the M\"obius amplitude ``knows'' about supersymmetry breaking, without however
 generating NS-NS tadpoles (of course, RR tadpoles are non-negotiable and always have to cancel). 
  
  The precursor paper of Angelantonj and Cardella contains a  model equivalent to the one presented above \cite{Angelantonj:2004cm}.\footnote{In the model constructed in Sect. 3 of  \cite{Angelantonj:2004cm}, the $\overbar{ {\rm O}7}_+$-$\overbar{ {\rm O}7}_-$ pair sits on the diagonal of the two-torus, which corresponds to a different choice of the projector in the M\"obius, $\Pi_{\rm RR} = (-1)^{m_9+m_8}  \Pi_{\rm NSNS}$. The 8 D7-branes were separated into two stacks of four branes, sitting on top of the anti-orientifolds. The attraction of the D7-branes on top of $\overbar{ {\rm O}7}_+$ cancels the repulsion of the D7-branes on top of $\overbar{ {\rm O}7}_-$. However, as already known by the authors of \cite{Angelantonj:2004cm}  and obvious from the discussion below, this configuration is unstable, since the D7-branes on top of  $\overbar{ {\rm O}7}_-$ are repelled by $\overbar{ {\rm O}7}_-$ and attracted towards the $\overbar{ {\rm O}7}_+$ -plane, leading to the stable configuration with one stack of eight coincident D7-branes, negative potential  and $\text{USp}(16)$ gauge group discussed above.} In our work,  we provide a microscopic geometrical interpretation of the source of supersymmetry breaking in terms of the replacement of an O$7_+$-O$7_-$ pair  by an $\overbar{ {\rm O}7}_+$-$\overbar{ {\rm O}7}_-$ pair, which leads to a novel form of brane supersymmetry breaking without tadpoles. Moreover, as argued in the next subsection, we believe that  the closed-string sector is not supersymmetric at tree-level but features spontaneously broken supersymmetry. 
  
 Let us now make considerations of energetics. The O-planes have no dynamical positions. The D7-branes, on the other hand, do have dynamical positions. To find which configuration is expected to minimize the one-loop effective potential, recall that the D7-branes are mutually BPS and have therefore no net interactions with the O${7_+}$ and O${7_-}$-planes. On the contrary, they are  attracted  by the  $\overbar{ {\rm O}7}_+$-plane and repelled by the $\overbar{ {\rm O}7}_-$-plane. Hence, the only stable configuration is obtained by putting all D7-branes on top of the 
 $\overbar{ {\rm O}7}_+$-plane, leading to a $\text{USp}(16)$ gauge group and breaking SUSY at the string scale, as explained above.   At first sight, one could think that a second option would be to put some stuck (or rigid) half-D7-branes on top of   O${7_-}$ or $\overbar{ {\rm O}7}_-$-planes, with no gauge group (but a $\mathbb{Z}_2$ global symmetry). As will be seen in Sect.~\ref{sec:consistency}, such configurations are however inconsistent, a fact that can be checked by adding probe D3-branes. 
 
To confirm these expectations, we write the M\"obius amplitude for arbitrary brane positions along $X^9$ and $X^8$. To this end, we introduce vectors $\vec{a}_\alpha=(a_\alpha^8,a_\alpha^9)$  such that the position of the half-brane $\alpha\in\{1,\dots,16\}$ along direction $X^i$ is $2\pi a_\alpha^iR_i$. In both channels, we obtain
\begin{align}
\begin{split}
\M&=\half\sum_\alpha\int\frac{\dd\tau_2}{\tau_2^5}\, \Bigg\{\,\left[(-1)^{n_9}W_{2n_8+2a_\alpha^2}-W_{2n_8+1+2a_\alpha^2}\right]\frac{\hat V_8}{\hat \eta^8}\\
&\hspace{2.8cm}-\left[(-1)^{n_9}W_{2n_8+1+2a_\alpha^2}-W_{2n_8+2a_\alpha^2}\right]\frac{\hat S_8}{\hat \eta^8}\Bigg\}\,W_{n_9+2a_\alpha^1}\ ,\\
\tilde{\M}&=\frac{\alpha'}{2R_9R_8}\sum_{\alpha}\int\dd l \, e^{4i\pi m_9 a_\alpha^1\, }e^{2i\pi m_8a_\alpha^2}\, P_{m_8}\, \Bigg\{\,\left[e^{2i\pi a_\alpha^1}P_{2m_9+1}-(-1)^{m_8}P_{2m_9}\right]\frac{\hat V_8}{\hat \eta^8}\\
&\hspace{7.15cm}-\left[e^{2i\pi a_\alpha^1}(-1)^{m_8}P_{2m_9+1}-P_{2m_9}\right]\frac{\hat{S}_8}{\hat \eta^8}\Bigg\}\, .
\end{split}
\end{align}
The dependance of the one-loop effective potential $\V$ on the independent positions can be derived solely from $\M$ and $\tilde \M$ in various regimes of  the internal radii. Among the $16$ vectors $\vec a_\alpha$, at most $8$ are dynamical degrees of freedom since the half-branes move by pairs and unpaired  half-branes stuck at a fixed point are not dynamical. We will label the dynamical ones by an index $r$. In Fig.~\ref{vector_field}, we display the vector field $ \big(\!\!-\!\!\frac{\partial\V}{\partial a_r^8},-\frac{\partial\V}{\partial a_r^9} \big)$ obtained numerically for a given brane $r$ of arbitrary position. As anticipated before, the minimum of the potential is reached when the branes sit at the origin, on the $\overbar{ {\rm O}7}_+$-plane. 

\begin{figure}[h!]
\begin{center}
\includegraphics[scale=1]{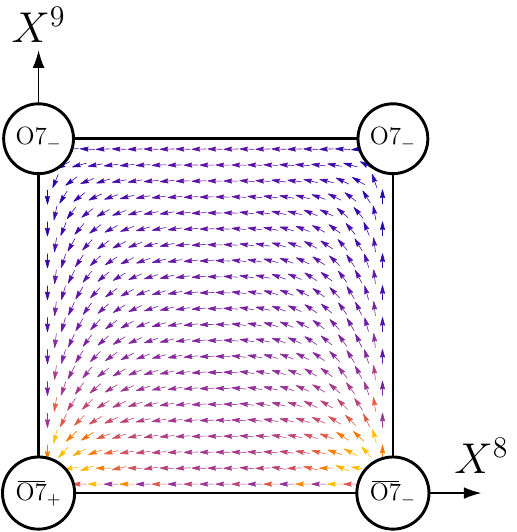}
\end{center}
\caption{ \footnotesize Example of vector field  $\big(\!\!-\!\!\frac{\partial\V}{\partial a_r^8},-\frac{\partial\V}{\partial a_r^9} \big)$ obtained numerically. The lighter the color is, the longer the vector norm  is.}
\label{vector_field}
\end{figure}

\subsection{Consistent pairing of torus and orientifold amplitudes}
\label{sec:non-susy-torus}

At first sight, one may think that the supersymmetric torus amplitude (\ref{8d1}) as well as the non-supersymmetic ones  (\ref{torus2}) and (\ref{torus3}) are all consistent   with the supersymmetric orientifold amplitudes (\ref{8d2}) and their non-supersymmetric deformation (\ref{8d12}). If true, this would yield six different orientifold models. 
There are however arguments based on the understanding of the orientifold projections as well as on the effective field theories that  suggest that only two options are consistent. 

Let us combine the non-supersymmetric torus amplitude (\ref{torus2}) with the non-super\-symmetric orientifold amplitudes (\ref{8d12}). 
The  torus amplitude can be constructed as an orbifold generated by $g' = (-1)^F \delta_{w_8} \delta_{p_9}$, while the orientifold amplitudes are obtained from the action of  $\Omega''= \Pi_8 \Pi_9 (-1)^{F_{\rm L}} (-\delta_{w_9})^F$. Hence, the (anti-)orientifold planes are located at the fixed points of $\Omega''$ and $\Omega''g'$. To be specific, $\Omega''$ fixes $(0,0)$ as well as $(\pi R_8,0)$, thanks to the $2\pi R_8$ periodicity. Moreover $\Omega''g'$ fixes $(0,\pi R_9)$ because $g'$ contains a factor $\delta_{p_9}$ acting as $X_9 \to X_9 + 2 \pi R_9$ after rescaling of the radius $R_9 \to 2 R_9$. It also fixes $(\pi R_8,\pi R_9)$ thanks to the $2\pi R_8$ periodicity. Notice that the presence of a factor $(-1)^F$ in such a group element changes the orientifolds into anti-orientifold planes, as can be seen in the M\"obius amplitude where $(-1)^F$ changes the sign of the RR coupling. As a result, the fixed points of $\Omega''$ are anti-orientifold planes, while those of $\Omega''g'$ are orientifold planes due to the cancellation of the factors  $(-1)^F$. Therefore, the nature of the O-planes derived from the non-supersymmetric amplitudes (\ref{8d12}) and shown in Fig. \ref{8d-square3} are in  agreement with the non-supersymmetric torus amplitude  (\ref{torus2}).

As shown in Sect. \ref{sec:8d1}, the second non-supersymmetric torus amplitude (\ref{torus3}) is equivalent to (\ref{torus2}) under the interchange of the coordinates $X^8 \leftrightarrow X^9$. Therefore, a  consistent orientifold model is obtained by applying  the same operation on the non-supersymmetric orientifold amplitudes (\ref{8d12}) and orientifold action (\ref{omega}). The corresponding geometry of O-planes is like in Fig. \ref{8d-square3}, with the ${\overline {\rm O7}_-}$-plane now located  at $(0,\pi R_9)$. 

Finally, reasoning as above with the generators ${g}=\delta_{w_8}\delta_{p_9}$  and $\Omega' = \Omega \Pi_8 \Pi_9 (-1)^{F_{\rm L}}$, one concludes that  the supersymmetric orientifold amplitudes (\ref{8d2}) are  compatible  with the supersymmetric torus amplitude (\ref{8d1}). 

In fact, the reason why the four other combinations of torus and orientifold amplitudes are inconsistent is that the orientifold projections are not symmetries of the closed-string spectrum. For instance, the supersymmetric torus amplitude (\ref{8d1}) does not seem to be compatible with the orientifold projection (\ref{omega}). Indeed, because the factor $(-\delta_{w_9})^F$ in $\Omega''$ is equivalent to $(-1)^{n_9F}$, fermions with even and odd winding numbers $n_9$ are projected differently. 
Due to interactions, the same conclusion should apply for bosons. 
However, there is no such selection rule in (\ref{8d1}), as opposed to the non-supersymmetric torus amplitude (\ref{torus2}), where fermions with even $n_9$ are gravitinos while those with odd $n_9$ are spin $\half$ particles.

Moreover, any attempt to combine the supersymmetric torus amplitude (\ref{8d1}) with the non-supersymmetric orientifold amplitudes (\ref{8d12}) is unlikely to be consistent, since the former implies the existence of massless gravitinos that would be difficult to explain from the point of view of the effective field theory. Instead, there seems to be no obstruction, from the point of view of the effective supergravity, to couple the non-supersymmetric torus amplitude (\ref{torus2}) with the orientifold ones (\ref{8d12}), since in this case all gravitinos are massive.

To understand a little better how the nature of (anti-)orientifold planes may or may not induce gravitino masses, let us consider first the orientifold projection $\Omega'$. In this case, ${\rm O}_{\pm}$-planes, which preserve the same supersymmetry, are located at the four fixed points of $T^2$. The action on the two  gravitinos $\psi_\mu$ and $\tilde\psi_\mu$ of the 10-dimensional type IIB is given by 
\begin{align}
&\Omega'\psi_\mu(X^8,X^9)(\Omega')^{-1}=\psi_\mu(-X^8,-X^9)=\Gamma_8\Gamma_9\tilde\psi_\mu(X^8,X^9)\ ,\nonumber \\
&\Omega'\tilde\psi_\mu(X^8,X^9)(\Omega')^{-1}=\tilde\psi_\mu(-X^8,-X^9)=-\Gamma_8\Gamma_9\psi_\mu(X^8,X^9)\ ,
\label{bc1}
\end{align}
which preserves one linear combination. 
Let us now consider  a geometry where anti-orientifold planes are located at the four fixed points. The action of the corresponding orientifold projection denoted $\tilde \Omega'$  preserves the orthogonal combination of gravitinos. Hence, we have  
\begin{align}
&\tilde\Omega'\psi_\mu(X^8,X^9)(\tilde\Omega')^{-1}=\psi_\mu(-X^8,-X^9)=-\Gamma_8\Gamma_9\tilde\psi_\mu(X^8,X^9)\ ,\nonumber \\
&\tilde\Omega'\tilde\psi_\mu(X^8,X^9)(\tilde\Omega')^{-1}=\tilde\psi_\mu(-X^8,-X^9)=\Gamma_8\Gamma_9\psi_\mu(X^8,X^9)\ .
\label{bc2}
\end{align}
We have seen that the geometry corresponding to the non-supersymmetric orientifold amplitudes (\ref{8d12}) involves both orientifold and anti-orientifold planes, as shown in Fig.~\ref{8d-square3}. Therefore, the boundary conditions of the gravitinos  at $X^9=0$ are of the  type (\ref{bc2}), whereas at $X^9 = \pi R_9$ they are of the type (\ref{bc1}). Overall, one obtains a shift in the KK spectrum of gravitinos $m_9/R_9 \to (m_9+\frac{1}{2})/R_9$, which is precisely what features the non-supersymmetric torus amplitude (\ref{torus2}) (see Eq. (\ref{grav-mass})). 

\section{Lower dimensional compactifications}
\label{sec:lower}

In this section, we extend the mechanism of supersymmetry breaking to models in even dimension $d\le 6$.  

In type IIB, a change of basis can always put an arbitrary  discrete background for the antisymmetric  tensor $B_{ij}$ into a  block-diagonal form,  with $2\times 2$ antisymmetric matrices,
\begin{equation}
B=\alpha'\begin{pmatrix}
0 & \lambda_d & & & \\
-\lambda_d & 0 & & (0) & \\
& & \ddots & & \\
& (0) & & 0 & \lambda_{9} \\
 & & & -\lambda_{9} & 0
\end{pmatrix}\ ,\quad \lambda_i\in\Big\{0,\half\Big\}\ . 
\end{equation}
The rank of the tensor $B_{ij}$ is twice the number of non-zero $\lambda_i$'s. Since they play no significant role in the sequel, we choose to set to zero the off-diagonal elements of the symmetric tensor $G_{ij}$. The internal space is thus a Cartesian product of circles of radii~$R_i$. 

In the supersymmetric case, one can switch on some $\lambda_i$'s by implementing a free-orbifold action on the background where $B_{ij}=0$.  For instance, $\lambda_9=\lambda_8=\half$ in 6d is achieved by considering the orbifold generated by  $g_1 = \delta_{w_8} \delta_{p_9}$ and $g_2 = \delta_{w_6} \delta_{p_7}$. In 4d, for $\lambda_7=\half$, one simply adds an extra generator $g_3 = \delta_{w_4} \delta_{p_5}$. 
By considering an orientifold action involving parities in all internal directions, one obtains a model  involving $2^{10-d}$ ${\rm O}(d-1)_\pm$-planes.  It is then  allowed to turn some ${\rm O}(d-1)_+$ - ${\rm O}(d-1)_-$ pairs into $\overbar{{\rm O}}(d-1)_+$ - $\overbar{{\rm O}}(d-1)_-$ ones in order to break supersymmetry. 

In the following, we consider various configurations of orientifold  planes of this type and provide the corresponding Klein, cylinder and M\"obius amplitudes. In orbifold language, the corresponding type IIB backgrounds  can be realized by  including operators $(-1)^F$ in the definition(s) of one or several of the generators $g_i$. 

\subsection{Geometry description}

The geometry of a model is given by the precise locations of the various ${\rm O}_\pm$ and $\overbar{{\rm O}}_\pm$-planes. Since pictorial representations become involved when the number of internal dimensions increases, it is useful to consider another way  to describe a generic geometry. If one specifies an ordering for the labelling of the fixed points, the geometry can simply be given by the list of O-plane types following this ordering. In the $10-d$ dimensional internal space, a fixed point can be represented by a $(10-d)$-vector with components $0$ or~$1$ that indicate if it is located at the origin or at $\pi R_i$ in each direction $X^i$, $i\in\{d,\dots,9\}$. For example, in 6d the fixed point located at $(X^6,X^7,X^8,X^9)=(0,\pi R_7,\pi R_8,0)$ is represented by $(0,1,1,0)$. 

In practice, let us label the fixed points by an index $A\in\{0,\dots, 2^{(10-d)}-1\}$. With this convention, their positions are given by $A$ written as a binary number. For instance in 6d, the fixed points are labelled as follows,
\begin{align}
\label{ord}
&A=0=(0,0,0,0)\ , &&A=1=(0,0,0,1)\ , && A=2=(0,0,1,0)\ ,\nonumber\\ &A=3=(0,0,1,1)\ ,&& ~~~~~~~~~~~~\cdots\,&& A=15=(1,1,1,1)\ .
\end{align}

\subsection{Models in six dimensions}

\noindent Projectors on the momenta in the Klein-bottle and M\"obius amplitude can either be factorized in the two internal $T^2$'s, or non-factorized. However, to obtain fully consistent models, compatibility of the projectors with the RR tadpole condition  turns out not to be sufficient. Indeed,   we give in the appendix examples of non-factorizable projectors, where one  is consistent and another one is not. In the following,  we will consider only consistent factorizable projectors. Supersymmetry breaking in the orientifold amplitudes will be implemented by choosing different projectors for the NS-NS and RR closed-string states propagating between the D-branes and the O-planes in the M\"obius amplitude. On the contrary, the Klein-bottle and cylinder amplitudes will take forms identical to those in the SUSY cases.

In six dimensions, a non-trivial antisymmetric tensor can have $\text{rank}\ B=2$ or $4$.\vspace{0.3cm}

$\bullet$ ${\rm rank} \ B = 2$: 8d model compactified on  $T^2$ (T-dualized)\vspace{0.2cm}

\noindent By compactifying the 8d model on an extra $T^2$ and T-dualizing both of its coordinates, one finds a 6d model with D5-branes and 16 O5-planes. In the supersymmetric case, one would have 12 O$5_-$ and  4 O$5_+$-planes. In the non-SUSY case, one obtains a configuration with 8 O$5_-$ and  $4 \times ( {\overbar{ {\rm O}5}_-} + {\overbar{ {\rm O}5}_+}) $-planes, where  the geometry is simply the 8d one duplicated along the new compact dimensions. The rank of the gauge group is $8$ and, depending on the location of the stacks of D5-branes, one finds for a single stack $\text{USp}(16)$ if the D5-branes are on top of one ${\overbar{ {\rm O}5}_+}$ or ${\rm O}5_+$-plane, $\text{SO}(16)$ if the D5-branes are on top of one ${\overbar{ {\rm O}5}_-}$-plane or O$5_-$-plane, or $\text{U}(8)$ if the stack of D5-branes is in the bulk. Supersymmetry is broken at the string scale (nonlinearly realized) if the D5-branes are coincident with anti-orientifolds, and broken only at the massive level (due to the separation in the internal space from the source of supersymmetry breaking) if the D5-branes are coincident with orientifold planes. 

When the D5-branes are put at the origin, the corresponding projectors on the momentum states running in the M\"obius amplitude  are
\begin{align}
 \Pi_{\rm NSNS} &= \frac{1-(-1)^{m_9}- (-1)^{m_8} - (-1)^{m_9+m_8}}{2}  \times   \prod_{i=6}^7 \frac{1+(-1)^{m_i}}{2}    \  ,  \nonumber \\
 \Pi_{\rm RR} &=  (-1)^{m_8}  \Pi_{\rm NSNS}     \ . \label{ldc1}  
\end{align}
The torus amplitude can be constructed as a free-orbifold generated by $g'_1 = (-1)^F \delta_{w_8} \delta_{p_9}$.
   
$\bullet$ ${\rm rank} \ B = 4$: \vspace{0.2cm}

 Following the ordering of the fixed points  given in Eq. ({\ref{ord}), the list of O$_\pm$-planes of the SUSY model we discuss here is
\begin{align}
&({\rm O}5_+,{\rm O}5_-, {\rm O}5_-,{\rm O}5_-, {\rm O}5_+,{\rm O}5_-, {\rm O}5_-,{\rm O}5_-, {\rm O}5_-,\nonumber \\
&\hspace{6cm}{\rm O}5_+, {\rm O}5_+,{\rm O}5_+, {\rm O}5_+,{\rm O}5_-, {\rm O}5_-,{\rm O}5_-)\ ,
\end{align}
with a total of 10 O$5_-$ and  6 O$5_+$-planes. There are several possible consistent SUSY breaking deformations. One example corresponds to a configuration containing 8 O$5_-$ and $4 \times ( {\overbar{ {\rm O}5}_-} +{\overbar{ {\rm O}5}_+})$-planes as follows, 
\begin{align}
&(\overbar{ {\rm O}5}_+,{\rm O}5_-,\overbar{ {\rm O}5}_-,{\rm O}5_-,\overbar{ {\rm O}5}_+,{\rm O}5_-,\overbar{ {\rm O}5}_-,{\rm O}5_-,\overbar{ {\rm O}5}_-,\nonumber\\
&\hspace{6cm}{\rm O}5_+,\overbar{ {\rm O}5}_+,{\rm O}5_+,\overbar{ {\rm O}5}_+,{\rm O}5_-,\overbar{ {\rm O}5}_-,{\rm O}5_-)\ .
\end{align}
The rank of the gauge group is $4$ and, depending
on the location of the stacks of D5-branes, one finds for a single stack $\text{USp}(8)$ if the D5-branes are on top of one ${\overbar{ {\rm O}5}_+}$ or O$5_+$ -plane, $\text{SO}(8)$ if the D5-branes are on top of one ${\overbar{ {\rm O}5}_-}$-plane or O$5_-$-plane, and $\text{U}(4)$ if the D5-brane stack is in the bulk. Supersymmetry breaking pattern is of course the same as in the ${\rm rank} \ B = 2$ case discussed above. 

When the 4 D5-branes (8 half-D5-branes) are coincident with the $\overbar{ {\rm O}5}_+$-plane at the origin of the internal space, the projectors on the momentum states in the tree-level channel M\"obius amplitude are 
\begin{align}
\Pi_{\rm NSNS} &= \frac{1-(-1)^{m_9}- (-1)^{m_8} - (-1)^{m_9+m_8}}{2}  \times     \frac{1+(-1)^{m_7}- (-1)^{m_6} + (-1)^{m_7+m_6}}{2}    \  ,  \nonumber \\
 \Pi_{\rm RR} &= (-1)^{m_8}  \Pi_{\rm NSNS}     \ . \label{ldc2}  
\end{align}
In order to write the orientifold amplitudes, it is convenient to denote lattices and volume factors as folows, 
\be
P_{\bf m}^{(10-d)}=\prod_{i=d}^9 P_{m_i} \ , \qquad W_{\bf n}^{(10-d)}=\prod_{i=d}^9 W_{n_i} \ ,\qquad v_{10-d}=\prod_{i=d}^9 R_i \ .
\ee
In these notations, the tree-level channel amplitudes are  given by  
\begin{align}
\mathcal{\tilde K} =&\; \frac{(\alpha')^2}{v_4 } \int_0^{\infty} \dd l \, P_{\bf m}^{(4)}  \frac{V_8-S_8}{\eta^8} \big(i l\big) \ , \nonumber \\
 \mathcal{\tilde A} =&\;\frac{2^{-5} N^2 (\alpha')^2}{2 v_4 } \int_0^{\infty} \dd l \, P_{\bf m}^{(4)}  \frac{V_8-S_8}{\eta^8} \big( il \big) \ , \nonumber \\
\mathcal{\tilde M} = &\, \frac{N (\alpha')^2}{4 v_4} \int_0^{\infty} \dd l \,  \big[  P_{2m_9+1} -   (-1)^{m_8} P_{2m_9}  \big]  P_{m_8} \big[  P_{2m_7} -   (-1)^{m_6} P_{2m_7+1}   \big]  P_{m_6}  \nonumber \\
& ~~~~~~~~~~~~~~~~~~~~\times \frac{{\hat V}_8- (-1)^{m_8}  {\hat S}_8}{{\hat \eta}^8} (i l+\tfrac{1}{2})   \  ,   \label{lcd3}  
\end{align}
while in the loop-channel they become
\begin{align}
\mathcal{K} = &\;\frac{1}{2} \int_0^{\infty} \frac{\dd \tau_2}{\tau_2^4} \, W_{2 {\bf n}}^{(4)}   \frac{V_8-S_8}{\eta^8} \big(2i \tau_2\big) \ , \nonumber \\
 \mathcal{A} =&\; \frac{N^2}{2} \int_0^{\infty} \frac{\dd \tau_2}{\tau_2^4} \,W_{\bf n}^{(4)}     \frac{V_8-S_8}{\eta^8} \big(\tfrac{i \tau_2}{2}\big)  \ , \nonumber \\
 \mathcal{M} =&\;  \frac{N}{2} \int_0^{\infty} \frac{\dd \tau_2}{\tau_2^4}  \, W_{n_9} \big[W_{2n_8} -(-1)^{n_9} W_{2n_8+1} \big]  W_{n_7} \big[   W_{2n_6} -  (-1)^{n_7} W_{2n_6+1} \big]  \nonumber \\
&~~~~~~~~~~~~~~~\, \times  \frac{  (-1)^{n_9} {\hat V}_8 + {\hat S}_8 }{{\hat \eta}^8} (\tfrac{i\tau_2}{2}+\tfrac{1}{2}) 
   \  .   \label{lcd4}  
 \end{align}
 The torus amplitude can be constructed as a free-orbifold generated by $g'_1 = (-1)^F \delta_{w_8} \delta_{p_9}$ and $g_2 =\delta_{w_6} \delta_{p_7}$. The presence of the factor $(-1)^F$ in $g'_1$ can be understood from the difference between the NS-NS and RR projectors in Eq.~(\ref{ldc2}), which is the same as in 8d examples. Actually, the same will be true  for all models we construct in what follows, only $g'_1$ will contain the supersymmetry breaking deformation $(-1)^F$.  

\subsection{Models in four dimensions}
\label{four_dimensions}

In four dimensions, a non-trivial antisymmetric tensor can have $\text{rank}\ B=2$, $4$ or $6$.\vspace{0.3cm}

$\bullet$  ${\rm rank} \ B = 2$: 8d model compactified on $T^4$ (T-dualized) \vspace{0.2cm}

\noindent By compactifying  the 8d model on $T^4$ and T-dualizing the four extra compact directions, one finds a 4d model with D3-branes and 64 O3-planes. In the supersymmetric case, one has $48 \ {\rm O}3_-$ and  $16 \ {\rm O}3_+$-planes. In the non-SUSY case, one obtains a configuration with $32 \ {\rm O}3_-$ and  $16 \times ( {\overbar{ {\rm O}3}_-} + {\overbar{ {\rm O}3}_+})$-planes. Like in the six dimensional ${\rm rank} \ B = 2$ case, the geometry is simply the one of the 8d model duplicated along the new compact directions. Since the model is T-dual to the 8d model compactified on an extra $T^4$, the rank of the gauge group is 8. For a single stack of D3-branes, the gauge symmetry is $\text{USp}(16)$ if the D3-branes are on top of one ${\overbar{ {\rm O}3}_+}$ or O$3_+$-plane, $\text{SO}(16)$ if the D3-branes are on top of one ${\overbar{ {\rm O}3}_-}$-plane or O$3_-$-plane, and $\text{U}(8)$ if the D3-brane stack is in the bulk. Supersymmetry is broken at the string scale (nonlinearly realized) if the D3-branes are coincident with anti-orientifolds, and broken only at the massive level (due to the separation in the internal space from the source of supersymmetry breaking) if the D3-branes are coincident with orientifold planes. 

When the D3-branes are put at the origin, the corresponding projectors on the momentum states running in the M\"obius amplitude are
\begin{align}
  \Pi_{\rm NSNS}& = \frac{1-(-1)^{m_9}- (-1)^{m_8} - (-1)^{m_9+m_8}}{2}  \times  \prod_{i=4}^7 \frac{1+(-1)^{m_i}}{2}   \  ,  \nonumber \\
\Pi_{\rm RR}& = (-1)^{m_8}  \Pi_{\rm NSNS}  \ . \label{ldc3}  
\end{align}

$\bullet$ ${\rm rank} \ B = 4$: 6d model compactified on $T^2$ (T-dualized)\vspace{0.2cm}

\noindent By compactifying  the 6d model on $T^2$ and T-dualizing the two extra compact directions, one finds a 4d model where, in the supersymmetric case, one has $40 \ {\rm O}3_-$ and  $24 \ {\rm O}3_+$-planes. In the non-SUSY case, one finds a configuration with $24 \ {\rm O}3_-$,   $8 \ {\rm O}3_+$  and  $16 \times ( {\overbar{ {\rm O}3}_-} + {\overbar{ {\rm O}3}_+}) $-planes. The geometry is the one of the 6d model duplicated along the two new dimensions. Since the model is T-dual to the 6d model compactified on $T^2$, the rank of the gauge group is $4$. For a single stack of D3-branes, the gauge symmetry is $\text{USp}(8)$ if the D3-branes are on top of one ${\overbar{ {\rm O}3}_+}$ or $ {\rm O}3_+$-plane, $\text{SO}(8)$ if the D3-branes are on top of one ${\overbar{ {\rm O}3}_-}$-plane or O$3_-$-plane, and $\text{U}(4)$ if the D3-brane stack is in the bulk. The supersymmetry breaking pattern is the same as in the previous cases. 

The corresponding projectors on the momentum states in the M\"obius for D3-branes put at the origin are
 \begin{align}
\Pi_{\rm NSNS} =&\;  \frac{1-(-1)^{m_9}- (-1)^{m_8} - (-1)^{m_9+m_8}}{2}  
  \nonumber \\
&\; \times \frac{1+(-1)^{m_7}- (-1)^{m_6} + (-1)^{m_7+m_6}}{2}  \times  \prod_{i=4}^5 \frac{1+(-1)^{m_i}}{2}   \  ,  \nonumber \\
 \Pi_{\rm RR} = & \;  (-1)^{m_8}  \Pi_{\rm NSNS}    \ . \label{ldc4}  
\end{align}

$\bullet$  ${\rm rank} \ B = 6$: \vspace{0.2cm}

\noindent Following our binary ordering, the geometry of the SUSY model discussed here is
\begin{align}
&\!\!( {\rm O}3_+,{\rm O}3_-, {\rm O}3_-,{\rm O}3_-, {\rm O}3_-,{\rm O}3_+, {\rm O}3_+,{\rm O}3_+, {\rm O}3_-,{\rm O}3_+, {\rm O}3_+,{\rm O}3_+, {\rm O}3_-,\nonumber\\
&{\rm O}3_+, {\rm O}3_+,{\rm O}3_+, {\rm O}3_-,{\rm O}3_+, {\rm O}3_+,{\rm O}3_+, {\rm O}3_+,{\rm O}3_-, {\rm O}3_-,{\rm O}3_-, {\rm O}3_+,{\rm O}3_-,\nonumber\\
& {\rm O}3_-,{\rm O}3_-, {\rm O}3_+,{\rm O}3_-, {\rm O}3_-,{\rm O}3_-, {\rm O}3_-,{\rm O}3_+, {\rm O}3_+,{\rm O}3_+, {\rm O}3_+,{\rm O}3_-, {\rm O}3_-,\nonumber\\
&{\rm O}3_-, {\rm O}3_+,{\rm O}3_-, {\rm O}3_-,{\rm O}3_-, {\rm O}3_+,{\rm O}3_-, {\rm O}3_-,{\rm O}3_-, {\rm O}3_-,{\rm O}3_+, {\rm O}3_+,{\rm O}3_+,\nonumber\\
& {\rm O}3_+,{\rm O}3_-, {\rm O}3_-,{\rm O}3_-, {\rm O}3_+,{\rm O}3_-, {\rm O}3_-,{\rm O}3_-, {\rm O}3_+,{\rm O}3_-, {\rm O}3_-,{\rm O}3_-)\ ,
\label{geom_6d_rk6susy}
\end{align}
with a total of $36$ O$3_-$ and  28 O$3_+$-planes. Again, there are several SUSY breaking deformations that are possible. One example  is a configuration with 20 O$3_-$, 12 O$3_+$ and   $16 \times ( { \overbar{{\rm O}3}_-} + { \overbar{{\rm O}3}_+}) $-planes as follows
\begin{align}
&(\overbar{ {\rm O}3}_+,{\rm O}3_-,\overbar{ {\rm O}3}_-,{\rm O}3_-,\overbar{ {\rm O}3}_-,{\rm O}3_+,\overbar{ {\rm O}3}_+,{\rm O}3_+,\overbar{ {\rm O}3}_-,{\rm O}3_+,\overbar{ {\rm O}3}_+,{\rm O}3_+,\overbar{ {\rm O}3}_-,\nonumber \\
&{\rm O}3_+,\overbar{ {\rm O}3}_+,{\rm O}3_+,\overbar{ {\rm O}3}_-,{\rm O}3_+,\overbar{ {\rm O}3}_+,{\rm O}3_+,\overbar{ {\rm O}3}_+,{\rm O}3_-,\overbar{ {\rm O}3}_-,{\rm O}3_-,\overbar{ {\rm O}3}_+,{\rm O}3_-,\nonumber\\
&\overbar{ {\rm O}3}_-,{\rm O}3_-,\overbar{ {\rm O}3}_+,{\rm O}3_-,\overbar{ {\rm O}3}_-,{\rm O}3_-,\overbar{ {\rm O}3}_-,{\rm O}3_+,\overbar{ {\rm O}3}_+,{\rm O}3_+,\overbar{ {\rm O}3}_+,{\rm O}3_-,\overbar{ {\rm O}3}_-,\nonumber\\
&{\rm O}3_-,\overbar{ {\rm O}3}_+,{\rm O}3_-,\overbar{ {\rm O}3}_-,{\rm O}3_-,\overbar{ {\rm O}3}_+,{\rm O}3_-,\overbar{ {\rm O}3}_-,{\rm O}3_-,\overbar{ {\rm O}3}_-,{\rm O}3_+,\overbar{ {\rm O}3}_+,{\rm O}3_+,\nonumber\\
&\overbar{ {\rm O}3}_+,{\rm O}3_-,\overbar{ {\rm O}3}_-,{\rm O}3_-,\overbar{ {\rm O}3}_+,{\rm O}3_-,\overbar{ {\rm O}3}_-,{\rm O}3_-,\overbar{ {\rm O}3}_+,{\rm O}3_-,\overbar{ {\rm O}3}_-,{\rm O}3_-)\ .
\label{geom_6d_rk6}
\end{align}
The rank of the gauge group is $2$ and, depending
on the location of the stacks of D3-branes, one finds for a single stack $\text{USp}(4)$ if the D3-branes are on top of one ${\overbar{ {\rm O}3}_+}$ or O$3_+$-plane, $\text{SO}(4)$ if the D3-branes are on top of one ${\overbar{ {\rm O}3}_-}$-plane or O$3_-$-plane, and $\text{U}(2)$ if the D3-brane stack is in the bulk.  

When the 2 D3-branes (4 half-D3-branes) are at the origin, the projectors on the momentum states in the M\"obius amplitude  are
\begin{align}
 \Pi_{\rm NSNS} &=  \prod_{i=2}^4  \frac{1-(-1)^{m_{2i+1}}- (-1)^{m_{2i}} - (-1)^{m_{2i+1}+m_{2i}}}{2}     \  ,  \nonumber \\
\Pi_{\rm RR} &= (-1)^{m_8}  \Pi_{\rm NSNS}    \ . \label{ldc5}  
\end{align}
The tree-level amplitudes are given by 
\begin{align}
 \mathcal{\tilde K} =&\;  \frac{(\alpha')^3}{4 v_6 } \int_0^{\infty} \dd l \, P_{\bf m}^{(6)}  \frac{V_8-S_8}{\eta^8} \big(i l\big) \ , \nonumber \\
   \mathcal{\tilde A} =&\; \frac{2^{-5} N^2 (\alpha')^3}{2 v_6 } \int_0^{\infty} \dd l \,  P_{\bf m}^{(6)}  \frac{V_8-S_8}{\eta^8} \big( il \big) \ , \nonumber \\
 \mathcal{\tilde M} = &- \frac{N (\alpha')^2}{8 v_4} \int_0^{\infty} \dd l \,  \prod_{i=2}^4 \big[  (-1)^{m_{2i}}   P_{2m_{2i+1}} -  P_{2m_{2i+1}+1}  \big]  P_{m_{2i}}   \nonumber \\
&~~~~~~~~~~~~~~~~~~~~~~~ \times \frac{{\hat V}_8- (-1)^{m_8}  {\hat S}_8}{{\hat \eta}^8} (i l+\tfrac{1}{2})   \  ,   \label{lcd6}  
\end{align}
while in the loop-channel they become
\begin{align}
 \mathcal{K} = &\;\frac{1}{2} \int_0^{\infty} \frac{\dd \tau_2}{\tau_2^3} \, W_{2 {\bf n}}^{(6)}   \frac{V_8-S_8}{\eta^8} \big(2i \tau_2\big) \ , \nonumber \\
  \mathcal{A} =&\;  \frac{N^2}{2} \int_0^{\infty} \frac{\dd \tau_2}{\tau_2^3}\, W_{\bf n}^{(6)}     \frac{V_8-S_8}{\eta^8} \big(\tfrac{i \tau_2}{2}\big)  \ , \nonumber \\
 \mathcal{M} = &\; \frac{N}{2} \int_0^{\infty} \frac{\dd \tau_2}{\tau_2^4} \,  \prod_{i=2}^4 W_{n_{2i+1}} \big[    (-1)^{n_{2i+1}} W_{2n_{2i}} -  W_{2n_{2i}+1} \big]     \, \frac{ {\hat V}_8 +  (-1)^{n_9}   {\hat S}_8 }{{\hat \eta}^8} (\tfrac{i\tau_2}{2}+\tfrac{1}{2}) 
   \  .   \label{lcd7}     
 \end{align}
  The torus amplitude can be constructed as a free-orbifold generated by $g'_1 = (-1)^F \delta_{w_8} \delta_{p_9}$, $g_2 =\delta_{w_6} \delta_{p_7}$ and $g_3 =\delta_{w_4} \delta_{p_5}$.

\section{Consistency conditions from probe branes}
\label{sec:consistency}

It is well-known that the standard consistency rules of  open-string partition functions are not enough to define a consistent string model. There are indeed K-theory constraints \cite{Minasian:1997mm,Witten:1998cd}, 
which  can also be understood in terms of the Witten $\text{SU}(2)$ anomalies  \cite{Witten:1982fp} on  probe branes \cite{Uranga:2000xp}. We are therefore interested in probe branes with $\text{SU}(2)$ gauge group. Probe branes mean  D-branes that are not constrained by the RR tadpole cancelation. In the type I string, background branes are D9 and the probe branes can be of D7, D5, D3 or D1 types, where  the D5 and D1-branes are BPS, whereas the D7 and D3 are non-BPS.  Since D1-branes lead to a 2d theory, whereas  we are interested 
in Witten four-dimensional $\text{SU}(2)$ anomalies, we will ignore D1-branes in what follows.  D7 and D3-branes in type I support unitary gauge groups $\text{U}(M)$, D5-branes support $\text{USp}(M)$ gauge groups  \cite{Sen:1999mg,Dudas:2001wd} for $B_{ij}=0$, whereas $\text{SO}(M)$ is also possible on D5-branes for $B_{ij}\not=0$.  
The cases of interest for us are SU$(2)\subset {\rm U}(2)$ and USp(2), which will be implicitly assumed in what follows.  
For Witten SU(2) anomaly, only the strings stretched between the background D9-branes and the probe branes, which transform in the fundamental representation of the SU(2) probe-brane gauge group, are relevant. Since the spectrum of these bifundamental strings are given entirely by the cylinder amplitude, supersymmetry breaking by the orientifold projection is not affecting our discussion below. In the following, we will first consider the corresponding spectra in type I language and  then perform T-dualities in all internal directions. After T-duality, one obtains a geometry with O$_+$,  O$_-$, $\overline{\rm O}_+$, $\overline{\rm O}_-$-planes, but due to the argument above one can restrict to configurations with O$_\pm$-planes only. The only cases giving constraints are when background branes are located on O$_-$-planes, which is implicitly assumed in what follows.

\subsection{Probe branes in eight dimensions}

In the type I string compactified to eight dimensions on $T^2$, two T-dualities switch the description into the type IIB$/\Omega'$ orientifold framework, where $\Omega'$ contains two parity operations $(X_8,X_9) \to (-X_8,-X_9)$.  One finds that: 

D9-D5 states contain six-dimensional Majorana--Weyl fermions in the $(M,2)$ of the gauge group $\text{SO}(M)_9 \times \text{USp}(2)_5$. After T-duality, if the D5 probe brane wraps  $T^2$, the configuration becomes D7-D3 with four-dimensional Weyl fermions in the bifundamental representation. Placing some D7 and D3 on the four O7-planes,  we learn that at each orientifold fixed point, $M$ should be even. 
Therefore stuck half-D7-branes {\it i.e.} without dynamical positions, and in particular $\SO 1$ configurations, are not allowed.  

Once this rule is satisfied, there are no further non-trivial constraints coming from D5 probe branes wrapped differently, or from    D7 and  D3 probe branes.

\subsection{Probe branes in six dimensions}

In type I string compactified to six-dimensions on $T^4$, the  probe branes of interest are D5-branes (which are not points in $T^4$), D7-branes whose worldvolume wrap $T^4$ and D3-branes. The strongest constraints come from D9-D5 strings, which become D5-D5 strings after four T-dualities, with the probe D5-branes wrapping a $T^2$ in the compact internal space.  
The consistent configurations with stacks of odd numbers of background half-D5-branes on  orientifold fixed points  must  contain  even numbers of such stacks in each $T^2$ in $T^4$.

\subsection{Probe branes in four dimensions}

In type I string compactified to four-dimensions on $T^6$, the only probe branes of interest are D7-branes and D3-branes. The former give no constraint since the number of D9-D7 Weyl fermions after reduction to four dimensions is always even. Moreover, after six T-dualities, the probe D3-branes become D9-branes which wrap the entire  internal space, while the background D9-branes become D3-branes. Hence, the probe D9-branes  intersect all background D3-branes, leading to no constraint.

\subsection{Extra non-perturbative constraint}

In addition to the Witten anomaly, another constraint on the allowed configurations comes from imposing that, in any dimension, the Wilson lines (or brane positions after T-duality) belong to $\text{SO}(N)$ (actually $\text{Spin}(N)$) and not $\text{O}(N)$. This is because at a nonperturbative level, the component of O$(N)$ disconnected from  $\SO N$ cannot be defined\cite{Witten:1998cd}. This implies that all  determinants of Wilson-line matrices must  equal one.

In practice, when the branes are located at fixed points, the matrix of Wilson lines along a direction $X^i$ is diagonal, with entries $1$ or $-1$ only. The number of $1$'s corresponds to the number of half-branes sitting at the origin of direction $X^i$ while the number of $-1$'s is the number of half-branes at position $X^i=\pi R_i$ \cite{ADM,ADM2,ADM3,ADM4,ADM5,ADM6,ADM7,ADM8,ADM9,ADM10,ADM11,ADM12,ADM13}. For the determinant to be one, we thus conclude that the allowed brane configurations are the ones for which the number of half-branes in each hyperplan $X^i=0$ or $X^i=\pi R_i$ is even.
 

\section{Gravitino ~mass ~\textit{versus} ~SUSY~breaking ~scale ~on \mbox{D-branes}}
\label{sec:massscales}

The main feature of the class of models constructed in this paper is the existence of two supersymmetry breaking mass scales: One in the closed-string sector, which is related to the compactification (KK or winding) scale, and another one in the open-string (gauge) sector, which is either the winding scale or the string scale, depending on which one is smaller. As mentioned in the introduction, this was already achieved in \cite{Angelantonj:2003hr}.  Our construction, which is motivated by the orientifold projection  put forward in the pioneering paper \cite{Angelantonj:2004cm}, allows one to avoid the open-string instabilities typically present in such constructions.
In the following, we  discuss in some details the mass scales and the limits where  supersymmetry is restored in the non-SUSY  8d model of Sect.~\ref{sec:8d}.

In the geometry of O-planes shown in Fig.~\ref{8d-square3},  the 8  D7-branes can be put in a single stack located in the bulk or coincident with one of the four orientifold planes. Let us consider the latter case. \vspace{0.3cm}

$\bullet$ Putting the stack on top of the ${\rm O}7_-$-plane closer to the ${\overline {\rm O7}_-}$-plane, the M\"obius contribution to  the vacuum energy is positive, since the D7-branes are repelled from the ${\overline{{\rm O}7}_-}$ plane and attracted towards the ${\overline{{\rm O}7}_+}$-plane. The M\"obius amplitude takes the form 
\begin{equation}
 \mathcal{M} =- \frac{N}{2} \int_0^{\infty} \frac{\dd \tau_2}{\tau_2^5}   \, W_{n_9} \big[ W_{2n_8} +  (-1)^{n_9} W_{2n_8+1} \big]  \frac{{\hat V}_8 - (-1)^{n_9}{\hat S}_8 }{{\hat \eta}^8} (\tfrac{i\tau_2}{2}+\tfrac{1}{2}) 
   \  ,   \label{ms1}
\end{equation}   
while the cylinder amplitude is still given by (\ref{8d12}). The ``field-theory'' open-string spectrum is encoded in 
\begin{align}
(\mathcal{A} +& \mathcal{M})|_{\rm FT} =\nonumber\\
&\int_0^{\infty} \frac{\dd \tau_2}{\tau_2^5}
   \left\{ \frac{N(N-1)}{2} \left.\left[
  W_{2n_9}W_{n_8} \frac{V_{8}-S_{8}}{\eta^8} 
   + W_{2n_9+1} \left( W_{2 n_8}  \frac{V_{8}}{\eta^8}
   - W_{2 n_8+1} \frac{S_{8}}{\eta^8} \right) \right]\right|_0 \right. \nonumber \\
&\left.\left. \hspace{1.4cm} + \frac{N(N+1)}{2} \, W_{2n_9+1} \left( 
 W_{2n_8+1 } \frac{V_{8}}{\eta^8} -
 W_{2n_8} \frac{S_{8}}{\eta^8}\right)\right|_0 \right\}    \,   , \label{ms2}
\end{align}
and is supersymmetric at the massless level. The gauge group is $\text{SO}(N)$, where \mbox{$N=16$} is fixed by the RR tadpole condition. Since the  closed-string spectrum becomes supersymmetric when $R_8 \to 0$ and/or $R_9 \to \infty$ (in particular gravitinos become massless), it is interesting to take these  limits in the open (gauge) spectrum. The first limit   $R_8 \to 0$ leads to a supersymmetric spectrum on the D7-branes. 
 Indeed, the winding towers of bosons and fermions collapse to the same  value. Supersymmetry is broken only at the massive winding level and for $R_8 \leq \sqrt{\alpha'}$ can therefore be considered as spontaneous at the field-theory massless level, after including quantum corrections.
 In the other limit  $R_9 \to \infty$, the open-string states featuring supersymmetry breaking become infinitely heavy and decouple at low energy.\footnote{However, none of the limits has  a purely perturbative orientifold description. In the first case  $R_8 \to 0$ the open-string spectrum does not have a 9d interpretation, whereas in the second case  $R_9 \to \infty$ there are local charges and tensions that generate a strong backreaction (local tadpoles are not cancelled). }\vspace{0.3cm}

$\bullet$ Putting all D7-branes on one stack coincident with the 
${\rm O}7_-$-plane closer to the $\overline {{\rm O}7}_+$-plane, the M\"obius contribution to the vacuum energy is negative. Otherwise there are no major differences. \vspace{0.3cm}

$\bullet$ Let us now consider the case where all D7-branes are coincident with the  $\overline{{\rm O}7}_+$-plane, whose amplitudes are displayed in Eqs.~(\ref{8d12}) and (\ref{8d10}).
The ``field-theory'' open-string spectrum is encoded in 
\begin{align}
(\mathcal{A} +& \mathcal{M})|_{\rm FT} = \int_0^{\infty} \frac{\dd \tau_2}{\tau_2^5}
   \left\{ \frac{N(N+1)}{2} \,
   W_{2n_9} \left( W_{2 n_8}  \frac{V_{8}}{\eta^8}
   - W_{2 n_8+1} \frac{S_{8}}{\eta^8} \right) \right|_0  \nonumber \\
&\left.\left.  + \frac{N(N-1)}{2}  \left[ W_{2n_9}
\left( W_{2n_8+1 } \frac{V_{8}}{\eta^8} -
 W_{2n_8} \frac{S_{8}}{\eta^8}\right)  + W_{2n_9+1}W_{n_8} \frac{V_{8}-S_{8}}{\eta^8} \right]\right|_0 \right\}    \,  . \label{ms3}
\end{align}
The gauge group is $\text{USp}(N)$, where $N=16$ is fixed by the RR tadpole condition.
In this case, the pattern of supersymmetry breaking depends on the value
of $R_8$. If this radius is large (and in general when branes are coincident with anti-orientifold planes)  the breaking is at the string scale, with nonlinearly realized supersymmetry. This interpretation is valid in the regime of large $R_8$ and $R_9 > \sqrt{\alpha'}$, when there are light gravitinos in the spectrum.

In the limit \mbox{$R_8 \to 0$}, the spectrum becomes however supersymmetric. In fact, when $R_8$ is small,  supersymmetry can be interpreted as spontaneously broken, since there is a shift in the fermion masses compared to the bosons in the winding sector.  Although this seems similar to the
familiar Scherk--Schwarz breaking, the mechanism has also features  in common with Brane Supersymmetry Breaking since the deformation does not affect the cylinder but acts in the  M\"obius amplitude by exchanging symmetric with antisymmetric gauge-group representations for fermions (compared to bosons) in the open-string  spectrum. Throughout the paper, we have used the terminology ``supersymmetry breaking at the string scale'' for this situation, in order to distinguish it with the case where the D-branes are on top of ${\rm O}_\pm$-planes.   

When $R_9 \to \infty$ at fixed $R_8$, the spectrum encoded in (\ref{ms3}) does not become supersymmetric, whereas the closed-string spectrum does. This is interesting since one may think that an exact Brane Supersymmetry Breaking Spectrum is realized in this limit. If true, 
this would also be a counter-example of the gravitino mass conjecture put forward recently in 
\cite{Cribiori:2021gbf,Castellano}. However, when $R_9 \to \infty$, the local sources from D-branes and O-planes generate local tadpoles and thus large backreactions responsible for the breaking of the effective field theory description, as conjectured in \cite{Cribiori:2021gbf,Castellano}.\footnote{
Note however  that these references do not claim that a limit of zero gravitino mass is not possible. There are known examples of Scherk--Schwarz type where the whole spectrum and interactions (closed and open strings in orientifolds, or only closed strings in heterotic and type II strings) become supersymmetric in the decompactification limit. The claim is that such a  limit is not possible, within an effective field theory description, if there is some sector breaking supersymmetry in the limit of vanishing gravitino mass.} 
 On the other hand, the model shows that the value of the gravitino mass  $m_{3/2}$ can be decoupled from the size of the scalar potential $V$, for fixed values of the moduli. In particular, $|m_{3/2}| \ll |V|^{1/d}$ is possible. Hence, we do not see any fundamental reason in quantum gravity to necessarily have a high gravitino mass compared to the Hubble scale \cite{Dudas:2021njv}, as recently proposed in \cite{Kolb:2021xfn,Kolb2}.  \vspace{0.3cm}

$\bullet$ Finally, putting all D7-branes on one stack coincident with the $\overline{{\rm O}7}_-$-plane, the M\"obius contribution  to the vacuum energy is positive and the pattern of supersymmetry breaking is similar to that found in the previous case.


\section{Conclusions and open questions}
\label{sec:conclusion}

We have constructed supersymmetry-breaking orientifold models where a certain number  $n$ of O$_-$ (negative tension, negative charge) - O$_+$ (positive tension, positive charge) orientifold-plane pairs are transformed into  $n$ ${\overbar {\rm O}_-}$ (negative tension, positive charge) - ${\overbar {\rm O}_+}$ (positive tension, negative charge) pairs. The  anti-orientifold plane pairs preserve the other half of supersymmetries, compared to the other ingredients of the background, which are O$_\pm$-planes  and D-branes.  In the open-string sector, supersymmetry is only broken in the M\"obius vacuum amplitude, whereas the closed-string sector has softly broken supersymmetry. 

The main interest of this mechanism is that both the original O$_-$ - O$_+$ pairs and their SUSY breaking cousins  ${\overbar {\rm O}_-}$ - ${\overbar {\rm O}_+}$  have zero 
tension and charge, so that the total tension and charge of the models are unchanged upon replacement. Therefore there are neither NS-NS nor RR tadpoles generated in the process.
 
Depending on where the background D-branes sit in the internal space, their massless spectrum 
is supersymmetric or not. In the latter case, which corresponds to D-branes located on anti-orientifold planes, the pattern of supersymmetry breaking depends on the value of a radius. If it is large, the breaking is at the string scale, with nonlinearly realized supersymmetry. On the contrary, 
if the radius is small, the same configuration describes a spontaneous breaking of supersymmetry. In 
this regime, the winding states in the D-brane spectrum are light and  supersymmetry breaking can be interpreted as a shift in the fermion masses compared to the bosons in the winding sector. 
This seems similar to the more familiar breaking by compactification (Scherk--Schwarz), but it differs in detail in that the brane-brane cylinder amplitude is not subject to this shift.

Constructions  of this type naturally stabilize open-string moduli. Indeed,  energetic considerations favor the D-branes to sit
on top of ${\overline{\rm O}_+}$-planes, where the scale of supersymmetry breaking on their worldvolume is  maximal.

An interesting issue in such models is their effective field-theory limits. At first sight, as initially considered in  \cite{Angelantonj:2004cm}, it seems  possible that the closed-string spectrum is supersymmetric at tree level. However, we have provided arguments showing that this is not plausible, as it would be at odds with the boundary conditions of the gravitinos imposed by the simultaneous presence of orientifold and anti-orientifold planes, which suggests massive gravitinos.
 In fact, we have given reasons  in favor of a specific soft supersymmetry-breaking deformation in the closed-string sector, rendering massive the gravitinos. Moreover,  if an exact supersymmetric closed-string spectrum would be compatible with the orientifold amplitudes discussed in our paper, one would  obtain  new models of Brane Supersymmetry Breaking type. This would also give counter-examples of the gravitino mass conjecture  \cite{Cribiori:2021gbf,Castellano}, whereas the models constructed in our paper are in agreement with it. More generally, existence of models with exact supersymmetry in the closed-string sector and broken supersymmetry in the open-string  sector would contradict the conjecture in \cite{Cribiori:2021gbf,Castellano}.
Three more comments are in order here:

First of all, all known models of this type, which are of BSB type or  with internal magnetic fields and broken supersymmetry, have NS-NS disk tadpoles that  could plausibly  trigger a breakdown of the  effective field theory in the perturbative vacuum, in agreement with \cite{Cribiori:2021gbf,Castellano}. Secondly, the exact supersymmetry in the closed-string  sectors of such models is valid only at tree-level, as it is broken by quantum corrections induced by the open-string sector. It is unclear to us if the conjecture on the gravitino mass should apply to the classical theory (tree-level spectrum) or to the quantum one. Lastly, in the string models we have constructed, it is possible to decouple the gravitino mass from the size of the scalar potential. In particular, the gravitino can  be much lighter than the scale determined by the magnitude of the quantum potential. This fact  could play a role in inflationary models of the type studied in \cite{Ema:2016oxl,Terada,Kolb:2021xfn,Kolb2,Dudas:2021njv,Antoniadis:2021jtg}.


Eventually, the string models we have considered  are based on toroidal compactifications and the fermionic spectrum, once reduced to four dimensions, is non-chiral.  
It would be of course very interesting to construct chiral four-dimensional models with supersymmetry breaking, by combining the mechanism put forward in this paper 
with other ingredients producing chirality, like orbifolds and/or fluxes.  We hope to come back to this interesting question in the near future.


\section*{Acknowledgments}
E.D. thanks Carlo Angelantonj, Jihad Mourad, Gianfranco Pradisi and Augusto Sagnotti for valuable discussions and correspondence. E.D. was supported in part by the ``Agence Nationale de
la Recherche'' (ANR). 

\begin{appendices}
\makeatletter
\DeclareRobustCommand{\@seccntformat}[1]{%
  \def\temp@@a{#1}%
  \def\temp@@b{section}%
  \ifx\temp@@a\temp@@b
  \appendixname\ \thesection:\quad%
  \else
  \csname the#1\endcsname\quad%
  \fi
} 
\makeatother

\section{Consistency of supersymmetric  M\"obius projectors}
\renewcommand{\theequation}{A.\arabic{equation}}
\setcounter{equation}{0}

In the construction of models in even dimension lower than 8, we have chosen factorized M\"obius projectors for simplicity. However, this is not imposed by the  RR tadpole condition. In this appendix, we will confirm that non-factorized projectors can be fully consistent, but we will also see that imposing the RR tadpole condition is not enough to obtain a M\"obius projector fully consistent. In the latter case, the inconsistency can only be seen in the direct Klein-bottle amplitude, while it is invisible in the open-string sector. We will study supersymmetric examples in 6d with $\text{rank}\ B=4$.

\subsection{A consistent non-factorized M\"obius projector}

In 6d, with all D-branes located at the origin, a generic M\"obius projector contains $16$ terms, one for each fixed point with  appropriate phases. The RR tadpole condition constrains the overall charge of the O-planes and thus the number of ${\rm O}_-$ and ${\rm O}_+$-planes. In the projector, this translates into a given number of minus and plus signs. In 6d with maximal rank for $B_{ij}$, there are $10$ O$5_-$ and $6$ O$5_+$-planes. This yields $10$ terms with a minus sign and $6$ terms with a plus sign in the projector. The total number of possibilities that fulfils this requirement is ${16\choose 6}=8008$.

Let us look at the following projector,
\begin{align}
\Pi=\ &1+(-1)^{m_9}+(-1)^{m_8}-(-1)^{m_9+m_8}+(-1)^{m_7}-(-1)^{m_9+m_7}-(-1)^{m_8+m_7}\nonumber\\
&-(-1)^{m_9+m_8+m_7}+(-1)^{m_6}-(-1)^{m_9+m_6}-(-1)^{m_8+m_6}-(-1)^{m_9+m_8+m_6}\nonumber\\
&-(-1)^{m_7+m_6}-(-1)^{m_9+m_7+m_6}-(-1)^{m_8+m_7+m_6}+(-1)^{m_9+m_8+m_7+m_6}\ .
\label{proj_ok}
\end{align}
With all D-branes at the origin, we deduce the geometry of the model,
\begin{align}
&({\rm O}5_+,{\rm O}5_+, {\rm O}5_+,{\rm O}5_-, {\rm O}5_+,{\rm O}5_-, {\rm O}5_-,{\rm O}5_-, {\rm O}5_+,\nonumber\\
&\hspace{6cm}{\rm O}5_-, {\rm O}5_-,{\rm O}5_-, {\rm O}5_-,{\rm O}5_-, {\rm O}5_-,{\rm O}5_+)\ .
\end{align}
It turns out to give no projection in the tree-level channel Klein-bottle amplitude, as was the case in Eq. (\ref{lcd3}). This is thus consistent with the supersymmetric torus amplitude generated by $g_1$ and $g_2$. The transverse cylinder amplitude is also the one obtained in Eq. (\ref{lcd3}),  while the M\"obius amplitudes, both in tree-level and loop  channels, are
\begin{align}
\tilde{\M}=&-\frac{N (\alpha')^2}{4 v_4} \int_0^{\infty} \dd l \,\Big[\left(P_{2m_9}+P_{2m_9+1}(-1)^{m_6}\right)\left(P_{2m_8}P_{2m_7+1}+P_{2m_8+1}P_{2m_7}\right)\nonumber\\
&+\left(P_{2m_9+1}-P_{2m_9}(-1)^{m_6}\right)\left(P_{2m_8}P_{2m_7}-P_{2m_8+1}P_{2m_7+1}\right)\Big]P_{m_6}\frac{{\hat V}_8-{\hat S}_8}{{\hat \eta}^8} (i l+\tfrac{1}{2})\nonumber\\
\M= &\int_0^{\infty} \frac{\dd \tau_2}{\tau_2^4}\Big[W_{n_9}\left(W_{2n_8}W_{2n_7}-W_{2n_8+1}W_{2n_7+1}\right)\left(W_{2n_6}+W_{2n_6+1}(-1)^{n_9}\right)\nonumber\\
&+\left(W_{2n_8}W_{2n_7+1}+W_{2n_8+1}W_{2n_7}\right)\left(W_{2n_6}(-1)^{n_9}-W_{2n_6+1}\right)\Big]\frac{{\hat V}_8 - {\hat S}_8 }{{\hat \eta}^8} (\tfrac{i\tau_2}{2}+\tfrac{1}{2}). 
\end{align}
The tree-level channel M\"obius amplitude contains all momentum states, just like the tree-level Klein bottle and cylinder, so that the factorization property of the amplitudes is satisfied. The loop-channel M\"obius amplitude is also consistent with the cylinder since it contains the contributions of the same states with signs. We conclude that the projector~(\ref{proj_ok}) yields a fully consistent model.

\subsection{An  inconsistent M\"obius projector}

Now consider the following projector, which has the correct number of signs to satisfy the RR tadpole condition,
\begin{align}
\Pi=\ &1+(-1)^{m_9}+(-1)^{m_8}+(-1)^{m_9+m_8}+(-1)^{m_7}+(-1)^{m_9+m_7}-(-1)^{m_8+m_7}\nonumber\\
&-(-1)^{m_9+m_8+m_7}-(-1)^{m_6}-(-1)^{m_9+m_6}-(-1)^{m_8+m_6}-(-1)^{m_9+m_8+m_6}\nonumber\\
&-(-1)^{m_7+m_6}-(-1)^{m_9+m_7+m_6}-(-1)^{m_8+m_7+m_6}-(-1)^{m_9+m_8+m_7+m_6}\ .
\label{proj_not_ok}
\end{align}
With all D-branes at the origin, the distribution of O-planes is given by  
\begin{align}
&({\rm O}5_+,{\rm O}5_+, {\rm O}5_+,{\rm O}5_+, {\rm O}5_+,{\rm O}5_+, {\rm O}5_-,{\rm O}5_-, {\rm O}5_-,\nonumber \\
&\hspace{6cm}{\rm O}5_-, {\rm O}5_-,{\rm O}5_-, {\rm O}5_-,{\rm O}5_-, {\rm O}5_-,{\rm O}5_-)\ .
\end{align}
With this geometry, the tree-level channel  Klein bottle is now different and not all momentum states are present. The tree-level and loop-channel Klein-bottle and M\"obius amplitudes are
\begin{align}
\tilde{\mathcal K} &=\; \frac{(\alpha')^2}{v_4 } \int_0^{\infty} \dd l \, P_{2m_9}\left(P_{m_8}P_{m_7}P_{m_6}+8P_{2m_8}P_{2m_7}P_{2m_6+1}\right)  \frac{V_8-S_8}{\eta^8} \big(i l\big) \ ,\nonumber \\
\mathcal K&=\;\frac{1}{4} \int_0^{\infty} \frac{\dd \tau_2}{\tau_2^4} \, W_{n_9}\left( W_{2n_8} W_{2n_7} W_{2n_6}+ W_{n_8} W_{n_7}(-1)^{n_6}W_{n_6}\right)   \frac{V_8-S_8}{\eta^8} \big(2i \tau_2\big) \ ,\nonumber\\
\tilde{\mathcal M}&=\, \frac{N (\alpha')^2}{4 v_4} \int_0^{\infty} \dd l  \, P_{2m_9} \Big\{P_{2m_8}\left[P_{m_7}P_{m_6}-2P_{2m_7}\left(P_{2m_6}-P_{2m_6+1}\right)\right]\nonumber\\
&\hspace{4cm}+P_{2m_8+1}\left(P_{2m_7}-P_{2m_7+1}\right)P_{m_6} \Big\} \frac{{\hat V}_8- {\hat S}_8}{{\hat \eta}^8} (i l+\tfrac{1}{2})\ ,\nonumber \\
\mathcal{M}&=\frac{N}{2} \int_0^{\infty} \frac{\dd \tau_2}{\tau_2^4}\,   W_{n_9}W_{n_8}\left(W_{2n_7}W_{2n_6}-W_{n_7}W_{2n_6+1}+(-1)^{n_8}W_{2n_7+1}W_{2n_6}\right)\nonumber\\
&\hspace{9.2cm}\times \frac{{\hat V}_8 - {\hat S}_8 }{{\hat \eta}^8} (\tfrac{i\tau_2}{2}+\tfrac{1}{2})\ .
\end{align}
The loop-channel  M\"obius amplitude contains all states present in the cylinder. Moreover, the tree-level Klein-bottle, cylinder and M\"obius amplitudes respect amplitude factorization. The only inconsistency comes from the Klein bottle in the loop-channel, which contains states not present in the torus amplitude. This means that the RR tadpole condition is not enough to produce a consistent M\"obius projector. The inconsistency can only be seen in the closed-string sector and comes from the geometry of the model.

\end{appendices}


\end{document}